\documentclass[twocolumn,showpacs,prb,floatfix,amsmath,amssymb]{revtex4}

\usepackage{amssymb}
\usepackage{amsmath}
\usepackage{bm}
\usepackage{graphicx}
\usepackage{float}

\DeclareMathOperator{\Tr}{Tr} 

\DeclareMathOperator{\Max}{max}

\DeclareMathAlphabet{\mathpzc}{OT1}{pzc}{m}{it}
\newcommand{\imaginary}{i}

\begin{document}
\title{Quantum theory for electron spin decoherence induced by nuclear
  spin dynamics in semiconductor quantum computer architectures:
  Spectral diffusion of localized electron spins in the nuclear 
solid-state environment}
\author{W. M. Witzel} \affiliation{Condensed Matter Theory Center,
  Department of Physics, University of Maryland, College Park, MD
  20742-4111} \author{S. Das Sarma}
\affiliation{Condensed Matter Theory Center, Department of Physics,
  University of Maryland, College Park, Maryland 20742-4111} \date{\today}
\begin{abstract}
We consider the decoherence of a single localized electron spin due to
its coupling to the lattice nuclear spin bath in a semiconductor
quantum computer architecture.  In the presence of an external
magnetic field and at low temperatures, the dominant decoherence
mechanism is the spectral diffusion of the electron spin resonance
frequency due to the temporally fluctuating random magnetic field
associated with the dipolar interaction induced flip-flops of nuclear
spin pairs.  The electron spin dephasing due to this random magnetic
field depends intricately on the quantum dynamics of the nuclear spin
bath, making the coupled decoherence problem difficult to solve.  We
provide a formally exact solution of this non-Markovian quantum decoherence
problem which numerically calculates accurate spin decoherence at short times, 
which is of particular relevance in solid-state spin quantum computer
architectures.
A quantum cluster expansion method is developed, motivated, 
and tested for the problem of
localized electron spin decoherence due to dipolar fluctuations of lattice
nuclear spins.  The method is presented with
enough generality for possible application to other types of spin
decoherence problems.  We present numerical results which are in
quantitative agreement with electron spin echo measurements in 
phosphorus doped silicon.  We also
present spin echo decay results for quantum dots in GaAs which differ
qualitatively from that of the phosphorus doped silicon system.  Our
theoretical results provide the ultimate limit on the spin coherence
(at least, as characterized by Hahn spin echo measurements)
of localized electrons in semiconductors in the low temperature and
the moderate to high magnetic field regime of interest in scalable
semiconductor quantum computer architectures.

\end{abstract}
\maketitle

\section{Introduction}
Quantum computation considerations have generated a great deal of
recent interest in the old \cite{herzog56, feher59, 
mims60_61, klauder62, zhidomirov69, chiba72} problem of electron spin
coherence in semiconductors.  In particular, a localized electron spin
can act as a qubit (i.e., a quantum dynamical two-level system) for
quantum information processing in scalable solid-state quantum
computer architectures. \cite{kane98, loss98}
But using such spin qubits for quantum
information processing purposes necessarily requires long spin
coherence time, and therefore understanding electron spin decoherence
in the solid-state environment becomes a key issue.  In this paper,
we develop a quantum theory for, what we consider to be, the most
important electron spin decoherence mechanism in spin-qubit based
semiconductor quantum computer architectures.  The spin decoherence
mechanism we consider here is the so-called spectral diffusion
mechanism, which has a long history, \cite{herzog56, feher59, 
mims60_61, klauder62, zhidomirov69, chiba72} and has been
much-studied recently\cite{desousa03b, tyryshkinP, abe04, witzel_short,
yao05} in the context of spin qubit
decoherence.

To provide a physical background for the theory to be presented in
this paper we start by considering a localized electron in a solid,
for example, a donor-bound electron in a semiconductor as in the doped
Si:P system.  Such a Si:P system is the basis of the Kane quantum
computer architecture.\cite{kane98}  The electron spin could decohere
through a number of mechanisms.  In particular, spin relaxation would
occur via phonon or impurity scattering in the presence of spin-orbit
coupling, but these relaxation processes are strongly suppressed in
localized systems and can be arbitrarily reduced by lowering the
temperature.  In the dilute doping regime of interest in quantum
computation, where the localized electron spins are well-separated
spatially, direct magnetic dipolar interaction between the electrons
themselves is not an important dephasing mechanism.\cite{desousa03a}
Interaction between the electron spin and the nuclear spin bath is
therefore the important decoherence mechanism at low temperatures and
for localized electron spins.   Now we restrict ourselves to a
situation in the presence of an external magnetic field (which is the
situation of interest to us in this paper) and consider the spin
decoherence channels for the localized electron spin interacting with
the lattice nuclear spin bath.  Since the gyromagnetic ratios (and
hence the Zeeman energies) for the electron spin and the nuclear spins
are typically a factor of 2000 different (the electron Zeeman energy
being larger), hyperfine coupling induced direct spin-flip transitions
between electron and nuclear spins would be impossible at low
temperature since phonons would be required for energy conservation.
\cite{footnote1}
This leaves the
indirect spectral diffusion mechanism\cite{footnote2} as the most effective electron
spin decoherence mechanism at low temperatures and finite magnetic
fields.  
The spectral diffusion process is associated with the
dephasing of the electron spin resonance due to the temporally
fluctuating nuclear magnetic field at the localized electron site.
These temporal fluctuations cause the
electron spin resonance frequency to diffuse in the frequency space,
hence the name spectral diffusion.  
The specific electron spin spectral diffusion process being considered
in this work 
is that arising from
the magnetic dipolar interaction induced flip-flop mechanism in
nuclear spins.  Other local flip-flop mechanisms between
nuclei, such as indirect exchange interactions
\cite{shulman58, bloembergen55, sundfors69} (which we will briefly 
address in this paper in the context of GaAs), may be easily
included in our formalism.  

Spectral diffusion (SD) is, in principle, not a limiting decoherence
process for silicon or germanium based quantum computer architectures
because these can, in principle, be fabricated free of nuclear spins 
using isotopic purification. Unfortunately this is not true for the 
important class of materials based on III-V compounds, 
where SD has
been shown to play a major role.\cite{desousa03a,desousa03b}  There
is as yet no direct (e.g., GaAs quantum dots) experimental measurement
of localized spin dephasing in III-V materials, but such experimental 
results are anticipated in the near future.  Indirect spin echo
measurements based on singlet-triplet transitions in coupled GaAs
quantum dot systems\cite{petta05} give $T_2$ times consistent with our
theoretical results.

SD is a theoretically challenging problem because
the temporally fluctuating random field causing the electron spin
dephasing is a non-Markovian stochastic dynamical variable arising
from the complex dipolar quantum dynamics of a large number of
interacting nuclear spins.  Although SD is an
intrinsic dephasing mechanism contributing to the so-called $T_2$-type
(i.e. ``transverse'') spin decoherence in the phenomenological Bloch
equation language, the effect of SD on the electron
spin dynamics (for example, on the electron spin resonance
measurement) cannot be characterized by a simple relaxation (or
dephasing) time $T_2$ associated with a ``trivial'' $e^{-t/T_2}$ type of
temporal loss of coherence.  The non-Markovian nuclear spin bath
dynamics produces a rather complex electron spin dephasing which
cannot be characterized by a single decoherence time (except as a very
crude approximation).  It is therefore more appropriate to consider
SD in the context of a specific experimental
situation, and it has been traditional, going back to the seminal
pioneering work of Hahn, to study SD in the context of
the spin echo decay in pulsed spin resonance 
measurements.\cite{hahn50, herzog56}  
In spin echo measurements, the inhomogeneous broadening
effects are suppressed by design, and any amplitude decay in the
consecutive pulses arises entirely from the $T_2$-dephasing time
associated with SD (and, of course, other applicable
$T_2$-type transverse relaxation processes).  

It was realized a long time ago\cite{herzog56,klauder62} 
that SD due to the
dipolar fluctuations of nuclear spins often dominates the coherence
decay in electron spin echo experiments. 
All available theories to date are based on classical stochastic 
modeling of the nuclear field, a Markovian theoretical framework 
which is inevitably
phenomenological since it requires an arbitrary choice for the
spectrum of nuclear fluctuations.  Such a classical Markovian
modeling is arguably incompatible with the strict requirements of 
spin coherence and
control in a quantum information device. In addition, 
recent rapid experimental progress in
single spin measurements, \cite{elzerman04} which in the near future
promises sensitive measurements of quantum effects in spin resonance,
also warrants a quantum theory of SD.
In the current work, we
develop a quantum theory of spin echo decay in electron spin resonance
experiments due to the SD caused by the dipolar
interaction induced flip-flop nuclear spin dynamics.  We emphasize
that, in contrast to all the earlier theories of SD in
the literature, our theory is fully quantum mechanical and incorporates the
dipolar nuclear spin flip-flop dynamics microscopically without making
any phenomenological statistical (Markovian or otherwise)
approximations.
To the best of our knowledge, 
ours is the first fully quantum theory for electron spin SD 
in solid state systems.
A proper description of
coupled spin dynamics is rather difficult due to the absence of the
usual Wick's theorem for spin degrees of freedom. In that regard, 
variations of our
method may prove rather useful, since environmental spin baths are
ubiquitous in any device exploiting the coherent properties of quantum
spin systems.

Our theory produces an accurate quantitative and qualitative prediction 
of the Hahn echo decay.  
It was pointed out a long time ago
\cite{klauder62,chiba72,abe04} that the observed time dependence of
these echoes are well fitted to the expression $\exp{(-t^2)}$, 
a behavior which can be derived
phenomenologically\cite{klauder62} by assuming
Lorentzian Brownian motion for the electron spin Zeeman frequency.
In our method this approximate behavior arises naturally from
the collective quantum evolution of the dipolar coupled nuclei,
without any phenomenological assumption on the dynamics of the
environment responsible for decoherence.  
Our theory reveals that the inclusion of quantum corrections to
nuclear spin fluctuation increases the degree of decoherence at lower
to intermediate time scales, as is best evidenced from our explanation
of the existing factor of $3$ discrepancy between the Markovian stochastic
theory \cite{desousa03b} and experimental data
\cite{chiba72,tyryshkinP, abe04} of spin echo decay in phosphorus doped
silicon.   

The rest of this paper is organized as follows.  In
Sec.~\ref{background} 
we provide a brief background for the spectral
diffusion process; in Sec.~\ref{problem_formulation} we formulate
the theoretical problem of spectral diffusion in the context of Hahn
spin echo decay; in Sec.~\ref{cluster_method} we develop and
describe the quantum cluster expansion which is the important
theoretical result of our work; in Sec.~\ref{perturbation_theories}
we describe dual perturbation theories in terms of a
$\tau$-expansion and a dipolar perturbation expansion 
which together explain the convergence of our cluster expansion and 
provide independent verification of its validity; and finally in
Sec.~\ref{results} we apply our theory to obtain numerical
results.  We conclude in Sec.~\ref{conclusion}.  Five appendixes
provide some mathematical details and proofs.

\section{Background}
\label{background}

\begin{figure}
\includegraphics[width=3.4in]{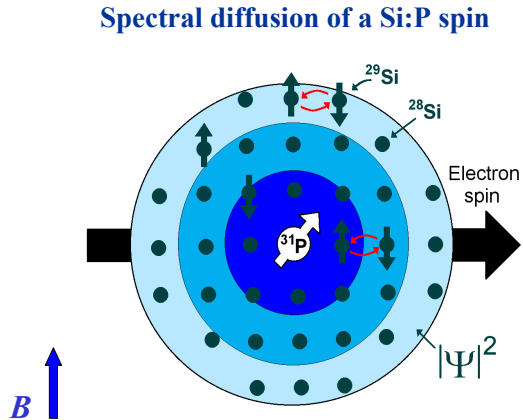}
\caption{
The electron of a P donor in Si experiences spectral
diffusion due to the spin dynamics of the eveloped bath of Si nuclei.
Of the naturally occuring isotopes of Si, only $^{29}$Si has a net
nuclear spin which may contribute to spectral diffusion by
flip-flopping with nearby $^{29}$Si.
Natural Si contains about $5\%$ $^{29}$Si or less through
isotopic purification.  Isotopic purification or nuclear polarization
will suppress spectral diffusion.
\label{SD_schematic}
}
\end{figure}

In systems of interest to us, a qubit is represented by the spin of a
localized electron, at low temperature and in a strong external
magnetic field, with a wave function that typically envelopes hundreds
of thousands of nuclei or more in the surrounding neighborhood of the lattice.
Two such examples are the shallow donor electron state 
in a phosphorus dopant in
silicon\cite{kane98} and a localized electron in a 
quantum dot in GaAs.\cite{loss98}
A schematic \cite{schematic} of the spectral diffusion process in 
the former is shown in Fig.~\ref{SD_schematic}.
As discussed in the Introduction, the dominant decoherence method 
is spectral diffusion (SD) caused by the
dynamic evolution of nuclear spins resulting in effective magnetic field
fluctuations experienced by the electron.
The electron spin couples to nuclear spins
via hyperfine interaction in the region in which the electron wave
function is appreciable.  The nuclear spins couple to each other via dipolar
interactions.  The large external field suppresses processes that do
not conserve electron or nuclear spin polarization in the direction of
the magnetic field.  Thus the only
relevant nuclear processes are flip-flops in which one spin is raised
while the other is lowered.

SD is a dephasing decoherence (i.e., a transverse or $T_2$-type
 relaxation) process, affecting only the precession
of the electron spin in the Bloch sphere without changing the
electron spin component along the magnetic field.  It thus contributes
to the energy-conserving $T_2$ decoherence rather than 
$T_1$ decoherence in which energy is exchanged with the bath
(Ref.~\onlinecite{hu01} details our definition of
$T_1$, $T_2$, and $T_2^*$).  The
$T_1$ time for these systems at low temperature is known to be much
longer than this $T_2$ time.  To analyze this decoherence, we consider
an ensemble of spins, initialized in some direction perpendicular to
the magnetic field, and observe the decay of the average spin over
time.
In experiments, many sparse donors
or dots serve as the ensemble, and they are initialized by applying a
$\pi/2$-pulse to rotate electron spins polarized along the magnetic field
axis into the plane perpendicular to the magnetic field.
For the free induction decay (FID), no further pulses are applied and
the observed decay is largely due to inhomogeneous broadening which is
caused by the difference in the local magnetic field of each electron,
causing precession at different rates for different electrons.
This will provide what is referred to as the $T_2^*$ decoherence time.
At the very least, this inhomogeneous broadening
is a result of the random initial distribution of nuclear spin states
which gives an uncertainty that scales as $\sqrt{N}$ where $N$ is the effective
number of nuclei ($\gtrsim 10^5$) 
influencing the electron significantly.  This
decoherence due to inhomogeneous broadening can be
eliminated by refocusing methods such as the Hahn echo.\cite{hahn50}
In such an experiment, a $\pi$-pulse, resonant with electron spins,
is applied at time $t = \tau$ in
the same direction as the original $\pi/2$ pulse and then an echo is
observed at $t = 2 \tau$.  This has the effect of reversing any static
local field effects and allows us to obtain the actual SD decoherence
(i.e., the $T_2$ decoherence as opposed to $T_2^*$ decoherence)
caused by effective magnetic field fluctuations.  This experiment
provides a measure of $T_2$.
The current paper will analyze the Hahn echo experiment.
More sophisticated pulses, such as the Carr-Purcell-Meiboom-Gill
sequence,\cite{meiboom58}
can yield even longer decoherence times and will be analyzed
in future publications.

Previous attempts at analyzing this SD decoherence have been based on
quasiclassical stochastic modeling.  Herzog and Hahn
\cite{herzog56} assigned a phenomenological Gaussian probability distribution 
function for the Zeeman frequency of the investigated spin without
considering the dynamics of the nuclear bath.  Later, Klauder and  
Anderson \cite{klauder62} used a Lorentzian distribution function
instead in order to account for a power law time dependence
observed in experiments by Mims and Nassau.\cite{mims60_61}
Zhidomirov and Salikhov \cite{zhidomirov69} devised a more 
sophisticated theory, with a wider range of
applicability, in which the flip rate
of each spin in the bath was characterized by Poisson distributions.  
Very recently de Sousa and Das Sarma,\cite{desousa03b}
in considering spectral diffusion by nuclear spin flip-flops,
extended this theory to characterize flip-flop rates of
pairs rather than individual spins within a phenomenological model.

In this paper, we present a microscopic theory that is based entirely
on the quantum mechanics of the system without resorting to phenomenological
distribution functions.  No Markovian assumption nor any assumption about the form of
the solution was used to obtain our results here.  Our expansion is
entirely based upon a density matrix formulation of the problem which assumes
infinite nuclear spin temperature ($T \gg$~nK) and uses
an approximate but microscopic Hamiltonian.  
The problem obviously involves too many
nuclear spins to solve directly using exact Hamiltonian diagonalization; 
however, the cluster expansion method we devise can give successive 
approximations to the
exact solution (convergent for short times, but often out to the tail of the
decay such that the full solution is obtained for practical purposes).  
A short report of our
results without any theoretical details has earlier appeared as a Rapid
Communication [\onlinecite{witzel_short}].
Our lowest order solution \cite{witzel_short} was 
recently reproduced by Yao {\it et al.}\cite{yao05} using an 
entirely different approach, thus providing an independent validation
of our theoretical approach.

\section{Formulation of the Problem}
\label{problem_formulation}

\subsection{Free evolution Hamiltonian}

The free evolution Hamiltonian for the spectral diffusion problem,
considering one localized electron spin, ${\bm S}$, and a nuclear spin bath,
${\bm I_n}$, in an external magnetic field $B$ (taken to be along the
$z$ direction), is given by
\begin{equation}
\label{H}
{\cal H} = {\cal H}^{Ze} + {\cal H}^{Zn} + {\cal H}^{A} + {\cal H}^{B},
\end{equation}
where the first two terms are due to the Zeeman energies of the
electron and nuclei respectively, ${\cal H}^{A}$ gives the
electron-nuclei coupling, and ${\cal H}^{B}$ contains the
internuclear coupling.

The Zeeman energy contributions, arising from the external magnetic
field, are given by
\begin{eqnarray}
\label{H_Ze}
{\cal H}^{Ze} & = & \gamma_{S} B S_{z}, \\
\label{H_Zn}
{\cal H}^{Zn} & = & -
B \sum_{n} \gamma_{n} I_{nz},
\end{eqnarray}
where $\gamma_{S}$ and $\gamma_{n}$ are the gyromagnetic ratios of the
electron spin and nuclear spins, respectively, and
$B$ is the external magnetic field defined to point in the $z$
direction.

The electron-nuclei coupling is given by
\begin{equation}
\label{H_A_exact}
{\cal H}^{A} = \sum_{n} A_{n} (I_{n} \cdot S).
\end{equation}
$A_n$ gives the magnetic coupling between the electron spin and each
nuclear spin.  It is dominated by the hyperfine coupling for nuclei
that contribute significantly to SD.
Since $\gamma_{S} = 1.76 \times 10^7 \mbox{(s G)$^{-1}$}$ is typically 
four orders of magnitude larger
than $\gamma_{n}$, flip-flop processes between the
electron and a nuclear spin are strongly suppressed at low temperatures
(i.e., no phonons) by energy conservation in a strong magnetic 
field.\cite{footnote1}
This direct hyperfine interaction leads to
quantitatively important effects at zero magnetic field,\cite{khaetskii02} but at
the moderate or high magnetic fields required for spin resonance measurements it only
contributes a small visibility decay.\cite{coish04}
We will therefore disregard the nonsecular part of the
electron-nuclei coupling, leaving us with
\begin{equation}
\label{H_A}
{\cal H}^{A} \approx \sum_{n} A_{n} I_{nz} S_{z},
\end{equation}
which we will use as the definition for ${\cal H}^{A}$ throughout the
rest of this paper.

The internuclear coupling due to the dipolar interaction is given by\cite{slichter}
\begin{eqnarray}
\label{H_B}
{\cal H}^{B} &=& \sum_{n \ne m} {\cal H}^{B}_{nm}, \\
\label{H_Bnm_exact}
{\cal H}^{B}_{nm} &=& \frac{\gamma_{n} \gamma_{m} \hbar}{2} 
\left[ \frac{I_{n} \cdot I_{m}}{R_{nm}^{3}} - \frac{3 (I_{n} \cdot
    {\bm R_{nm}}) (I_{m} \cdot {\bm R_{nm}})}{R_{nm}^5}\right],~~~
\end{eqnarray}
where ${\bm R_{nm}}$ is the vector joining nuclei $n$ and $m$.
This can be expanded into a form containing only operators of the type 
$I_{+}$, $I_{-}$, or $I_{z}$.\cite{slichter}  We will again invoke the
energy conservation argument which, because of the strong external
magnetic field, allows us to neglect any term that changes the total Zeeman energy 
of the nuclei.
This will leave us with the following secular contribution:
\begin{eqnarray}
\label{H_Bnm}
{\cal H}^{B}_{nm} &\approx& b_{nm}(
\delta{(\gamma_n - \gamma_m)}
I_{n+} I_{m-}  - 2 I_{nz} I _{mz}), \\
\label{bnm}
b_{nm}&=&-\frac{1}{4}\gamma_{n}\gamma_{m}\hbar\frac{1 - 3 \cos^{2}{\theta_{nm}}}{R^{3}_{nm}},
\end{eqnarray}
where $\theta_{nm}$ is the angle of ${\bm R_{nm}}$ 
relative to the magnetic field direction.  
Equation~(\ref{H_Bnm}) will be used for ${\cal H}^{B}_{nm}$ throughout the rest of
this paper.  Note that the flip-flop interaction between nuclei with 
different gyromagnetic ratios is suppressed by Zeeman energy
conservation in the same way that the nonsecular part of the dipolar 
interaction is suppressed.
This occurs, for example, in GaAs because the two
isotopes of Ga and the one isotope of As that are present have significantly
different gyromagnetic ratios.  This approximation is applicable if
any such differences in $\gamma_n$ are on the order of $|\gamma_n|$ itself
and thus such energy changes would be large compared to other energies
of the problem in the large external magnetic field being considered
in this work.  Giving some
typical numbers, $\Max{(|A_n|)} \sim 10^6~s^{-1}$, $\Max{(|b_{nm}|})
\sim 10^2~s^{-1}$, and for $B = 10~$T, $|\gamma_n B| \sim 10^8~s^{-1}$ 
and $|\gamma_{S} B| \sim 10^{12}~s^{-1}$.  The
Zeeman energies are thus much larger than the other energies in the problem.
In fact, our model of neglecting terms that change the total Zeeman
energy of the nuclei turns out to be reasonably valid for $B > 0.1~$T
or so for Si:P or GaAs systems of interest to us.  SD is
therefore the dominant spin decoherence channel for localized
electrons in semiconductors except at zero or very small ($< 0.1~$T)
magnetic fields.

\subsection{Hahn echo}

The Hahn echo experiment consists of preparing the electron spin in
a state perpendicular to the magnetic field, allowing free evolution
for a time $\tau$, applying an electron spin resonant $\pi$-pulse 
about an axis perpendicular to the magnetic field 
(for our theoretical purpose, the direction of
this axis is not important beyond being perpendicular to the
$B$-field), 
and then allowing free evolution again for time
$\tau$.  The echo envelope is the magnitude of the resulting ensemble
spin average at time $t = 2 \tau$.

Mathematically, the echo envelope is the magnitude 
of the expectation value of the spin multiplied by $2$ for
normalization (since the electron has a spin of $1/2$).  
Because we are only concerned with the component of
the spin perpindicular to the magnetic field, 
it is also the magnitude of the normalized complex
in-plane magnetization:
\begin{eqnarray}
\label{vE_physical}
v_{E}(\tau) &=& 2 (\langle S_x \rangle  + \imaginary
\langle S_y \rangle) \\
&=& 2~\Tr{\left\{(S_x + \imaginary
  S_y)\rho(\tau)\right\}}.
\label{amplitude_def}
\end{eqnarray}
The density matrix of the system, $\rho(\tau)$, is given by
\begin{equation}
\rho(\tau) = U(\tau) \rho_{0} U^{\dag}(\tau),
\label{rho_t} 
\end{equation}
with the evolution operator 
\begin{equation}
\label{U}
U(\tau)=e^{-\imaginary {\cal H} \tau}\sigma_{x, e}e^{-\imaginary {\cal H} \tau},
\end{equation}
where ${\cal H}$ is the free evolution Hamiltonian given by Eq.~(\ref{H}).
We have arbitrarily chosen the pulse axis to be in the $x$ direction.
This pulse is represented by the Pauli matrix, $\sigma_{x, e} = 2 S_{x}$ with
the $e$ subscript denoting that it is an electron spin operator.

Our initial density matrix, $\rho_0$, will be given by a product state
of the electron spin, in a direction perpendicular to the magnetic
field, and the nuclear spins in a thermal state:
\begin{eqnarray}
\label{rho_0}
\rho_{0} &=& |\chi_{e}^{0}\rangle\langle \chi_{e}^{0}| \otimes
\frac{e^{-{\cal H}^{n}/k_B T}}{M}, \\
\label{chi_0}
|\chi_{e}^{0}\rangle &=& \frac{1}{\sqrt{2}}(|0_e\rangle + e^{i \phi} |1_e\rangle),
\end{eqnarray}
where ${\cal H}^n={\cal H}^{Zn}+{\cal H}^B$ and $M$ is its partition
function ($M\approx 2^N$ for $T\gg $~nK, where $N$
is the number of nuclear spins).  The initial angle of the spin,
$\phi$, will contribute an irrelevant phase factor to the complex 
in-plane magnetization.
Dropping this irrelevant phase (since we are interested in the
magnitude), after a few manipulations, shown in Appendix
\ref{appendix_analysis}, Eq.~ (\ref{amplitude_def}) becomes
\begin{equation}
v_{E}(\tau)=\frac{1}{M} \Tr{\left\{
U_{-}U_{+}e^{-{\cal H}^n/k_B T}
U_{-}^{\dag}U_{+}^{\dag}
\right\}},
\label{v_E_temp}
\end{equation}
where 
\begin{equation}
U_{\pm}=e^{-\imaginary {\cal H}^{\pm}\tau}
\label{Upm}
\end{equation}
are evolution operators under the effective Hamiltonians
\begin{equation}
{\cal H}^{\pm}= \sum_{n \ne m} {\cal H}^{B}_{nm} \pm \frac{1}{2} \sum_{n} A_{n} I_{nz},
\label{Hpm}
\end{equation}
which describe nuclear evolution under the effect of an electron spin
up (${\cal H}^{+}$) or down (${\cal H}^{-}$) with the external
magnetic field dependence now canceled out.  Although the $U_{\pm}$
evolution operators are independent of $B$, it should be noted that a
strong external magnetic field is required to justify the approximations of
Eqs.~(\ref{H_A}) and (\ref{H_Bnm}).
The trace in
Eq.~(\ref{v_E_temp}) is taken over nuclear spin states only.

For our purposes we only consider high nuclear temperatures ($T \gg$~nK) 
such that we can make the following approximation:
\begin{equation}
\label{v_E}
v_{E}(\tau) = 
\frac{1}{M} \Tr{\left\{
U_{-}U_{+}U_{-}^{\dag}U_{+}^{\dag}
\right\}}.
\end{equation}
The high nuclear temperature (and the low electron and phonon
temperature) approximation made throughout this work is well-valid in
the solid-state quantum computer architectures of our interest, where
typical experiments would be carried out around $T \sim 100~$mK which
is an extremely high temperature for the nuclear bath dynamics and an
extremely low temperature for phonon excitations.

\section{Cluster Method}
\label{cluster_method}
Our cluster method provides a strategy to compute the Hahn echo, $v_{E}(\tau)$ 
[Eq.~(\ref{v_E})], in successive orders of accuracy which
may or may not converge for a specific application (naturally the
systems reported in this paper do converge, for practicle purposes, 
as we will show).  
We refer to these
successive orders, loosely, as an ``expansion'' although we do
not wish to suggest that corrections are additive (as in familiar
expansions such as a Taylor series).   
In Sec.~\ref{ConceptualClusterExpansion}, we provide
a conceptual description of this cluster expansion.
Section \ref{FormalClusterExpansion} 
supplies the mathematical formalism and practical implementation
of this expansion.

\subsection{Conceptual cluster expansion}
\label{ConceptualClusterExpansion}

\begin{figure}
\includegraphics[width=3in]{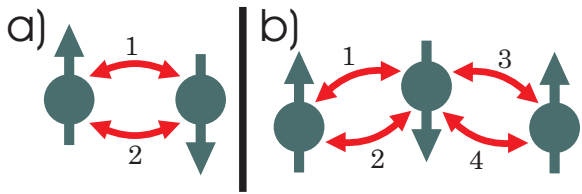}
\caption{
\label{nuclear_processes}
Some possible nuclear ``processes'' where the numbered
two-sided arrows represent a sequence of flip-flops between pairs of
nuclei.  (a) depicts a two-nuclei process and (b) depicts a three-nuclei
process.
}
\end{figure}

Consider independent, simultaneous nuclear ``processes'' that may
contribute to the decay of the Hahn echo.  For example, a process may
involve a pair of nuclei flip-flopping [Fig.~\ref{nuclear_processes}(a)]
which results in fluctuations of the
effective magnetic field seen by the electron spin, or it may involve 
three nuclei interdependently [Fig.~\ref{nuclear_processes}(b)], etc.
The dynamics of such a process results from the local coupling
between nuclei (with coupling constants $\{b_{nm}\}$), and hyperfine
coupling to the electron (with coupling constants $\{A_{n}\}$).
Any number of these processes may occur ``simultaneously'' as long as
they involve disjoint sets of nuclei and are thus independent of each
other (processes that share a nucleus are not independent and would
have to be combined into a larger process).

Using this (not yet well-defined) concept of nuclear processes,
the cluster expansion may be described, ideally, as follows.  
The cluster expansion
will include processes that involve a successively increasing
number of nuclei.  
Because an isolated nucleus in our model does not contribute 
to spectral diffusion, at the lowest nontrivial order 
we include any simultaneous
processes involving two nuclei (pairs). 
Thus at this lowest
order we can involve any number of pair processes together as long as
the pairs do not overlap (i.e., involve the same nucleus). 
At the next order, we will additionally include processes that involve three
nuclei.  Next, we include four nuclei processes which
cannot be decomposed into two pair processes (these would already have been
included).  
To summarize, let us say that the $k^{\mbox{\scriptsize th}}$ order of the
expansion will include processes of up to $k$ nuclei.

Because {\it all} processes involving a given number of nuclei are included
at each order of this expansion, and because these processes are
independent (proven formally in Sec.~\ref{FormalClusterExpansion}
and the related appendix), 
rather than working with individual processes,
we can work with contributions due to each given ``cluster'' of nuclei 
(for now, simply defined as a set of nuclei); such a ``cluster 
contribution'' includes contributions from all of the processes 
involving all nuclei in that cluster 
in an interdependent way (i.e., not separable into independent
sub-processes).
Thus we may say
that the $k^{\mbox{\scriptsize th}}$ order of the expansion includes
contributions from clusters up to size $k$.
These ``contributions'' are not necessarily additive in the
solution because we must account for simultanous but independent
processes (from disjoint clusters).  
The idea is simply to include the possibility
of interdependent processes involving clusters of successively increasing size.

\begin{figure}
\includegraphics[width=1.5in]{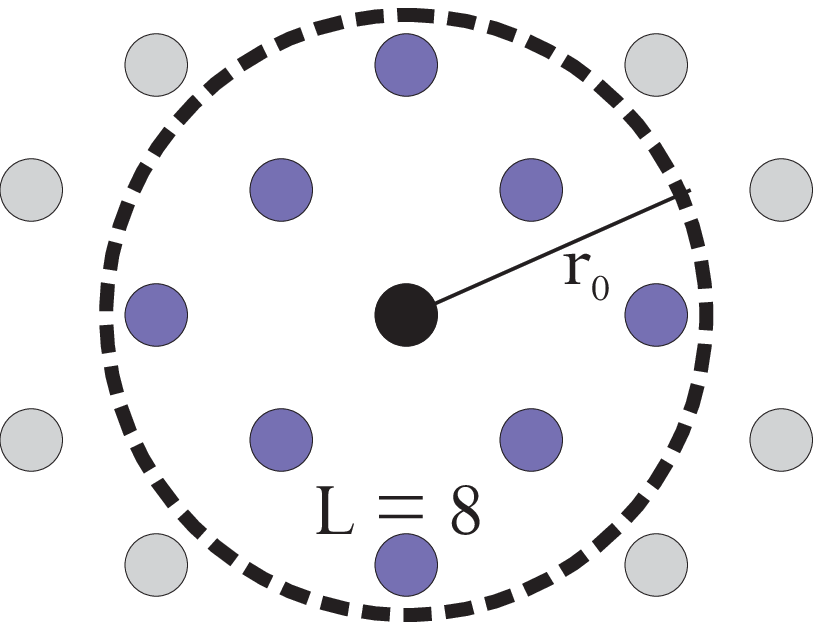}
\caption{
\label{Fig_L_from_r0}
$L$ is loosely defined as the average number of neighbors in a near
neighbor approximation that converges to the exact answer.
For example, we can include neighbors up to a distance 
$r_0$ away such that increasing this maximum neighbor distance in the
near neighbor approximation (where non-neighbor interactions are
neglected) does not significantly change the solution.
}
\end{figure}

We deliberately use the word ``cluster'' to imply proximity between
the members of the set of nuclei involved in interdependent processes.  
In fact, a near neighbor approximation, in which the constituent nuclei of a
contributing cluster must be in the same neighborhood, 
is justified by the $1 / {\bf R}^{3}_{nm}$ dependence of the
internuclear coupling constant [Eq.~(\ref{bnm})].
Consider a near (not necessarily nearest) neighbor approximation with
an adjustable parameter $r$ such that we ignore interactions 
between nuclei that are a further apart than $r$.  If a near
neighbor approximation is applicable, the Hahn echo solution in this
approximation (in principle, whether or not it is feasible to compute) 
will converge with an 
acceptable level of accuracy at some finite value of $r$ much 
smaller than the system size.  Let us define $r_0$ to be the value of
$r$ in which this acceptable convergence is achieved.  Let $L$ be the
number of nuclei within a range of $r_0$ from any nucleus, on
average, as shown in Fig.~\ref{Fig_L_from_r0}.  
Applying this near neighbor approximation to our cluster expansion,
$L$ determines the way in which the number of contributing clusters scales with
cluster size.  This has important implications for the convergence of
the cluster expansion.

\begin{figure}
\includegraphics[width=3.4in]{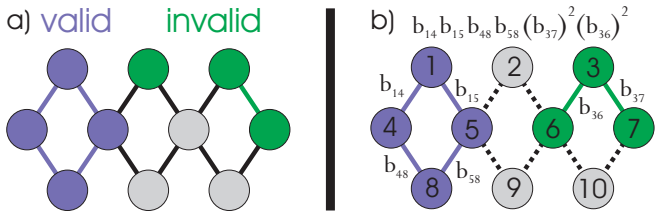}
\caption{
\label{FigConnectedCluster}
(Color online)
(a)~The set of nuclei in a contributing cluster must form a connected
graph.  Edges represent neighbor connections.  The set of blue
nuclei on the left form a valid cluster.  The set of green nuclei on
the upper right are not fully connected so they do 
{\it not} form a valid cluster.
(b)~A set of $b_{nm}$ factors in a term of an expansion of $v_E(\tau)$
determines a set of disjoint clusters (the connected subgraphs formed
from $b_{nm}$ edges).
}
\end{figure}

To be specific, the convergence of the cluster expansion depends upon
two factors.  The first is how the number of
contributing clusters scales with cluster size (which relates to $L$
as we have already said).  
The second is how the average contribution of clusters scales with 
cluster size.
Clusters only contribute via interdependent processes; thus the set 
of nuclei in a contributing cluster must form a connected graph where
edges in the graph connect neighbors [Fig.~\ref{FigConnectedCluster}(a)].
When counting the number of clusters of a given size,
we have $N$ sites to choose from for the first nucleus, 
but there are only ${\cal O}(L)$ possibilities (roughly) for each 
additional nucleus because it must neighbor one of the previous
choices.  This simple analysis does not compensate for overcounting
due to permuting labels and other such details, but it provides the
correct scaling in terms of $N$ and $L$; that is, there are
${\cal O}(N L^{k-1})$ contributing clusters of size $k$. 
Our first scaling factor is then
$L$ since the number of clusters as a function of cluster size scales
in powers of $L$.  The other scaling factor will rely upon 
some perturbation theory to describe how cluster contributions
themselves scale with an increase in cluster size.  Section
\ref{perturbation_theories} will discuss two perturbation
theories that may be applicable 
[for short times, $\tau \ll \Max{(b_{nm})}^{-1}$], 
but for now we will use $\lambda$ to represent some
perturbation parameter and assume that a cluster contribution of size
$k$ will scale as ${\cal O}(\lambda^k)$.  Thus $\lambda$ is assigned as 
the other scaling factor, and we may loosely argue that we expect the
cluster expansion to converge when $\lambda L \ll 1$ because the
total contribution from clusters then decreases as we increase in
cluster size.  This reasoning will become more rigorous 
when we go on to explain how we implement this cluster expansion.

\subsection{Mathematically formalized cluster expansion}
\label{FormalClusterExpansion}

In Sec.~\ref{ConceptualClusterExpansion}, we gave
a rough, conceptual description of our cluster
expansion to guide the reader's intuition and present some basic
ideas.  
At this point, we will
develop the rigorous mathematical formalism that relates the idea of
simultaneous, independent nuclear processes contributing to the
Hahn echo directly to our mathematical expression [Eq.~(\ref{v_E})] 
for the Hahn echo.  We will decompose Eq.~(\ref{v_E}) into a sum of
products of cluster contributions.  Each cluster contribution will
effectively contain the sum of contributions from all processes 
involving, inseparably (i.e., interdependently), all nuclei in the cluster.
Such a decomposition requires that processes involving 
disjoint sets of nuclei are truly independent
and interchangeable.  This requirement is met by proving, as we shall, that
a cluster contribution is independent of anything outside of the cluster.

These cluster contributions need not be computed by analyzing the
possible ``processes'' involving each set of nuclei.
Instead, the decomposition of Eq.~(\ref{v_E}) into cluster
contributions will be used recursively to define these cluster
contributions; this is shown in Sec.~\ref{clusterdefinition}.
With these cluster contributions
concretely defined, we then discuss, in
Sec.~\ref{IdealClusterExpansion}, 
how we mathematically define the 
ideal cluster expansion that we have conceptually
described.  This ideal expansion is useful for understanding some
basic ideas, but in order to practically perform calculations on large
systems, some further approximation techniques must be used.  This
practical implementation of the cluster expansion is explained in
Sec.~\ref{PracticalClusterExpansion}.

\subsubsection{Decomposing into cluster contributions}
\label{clusterdefinition}
Consider expanding $v_E(\tau)$ [Eq.~(\ref{v_E})] into a sum
of products with respect to internuclear coupling, 
having all $b_{nm}$ constants factorable from each term.
For example, such an expansion could
be made by Taylor expanding the exponentials of $U_{\pm}$
[Eq.~(\ref{Upm})] and then distributing through these sums.
Each term in such an infinite expansion has a set of 
$b_{nm}$ factors which determine a set of involved clusters.  
In the language of graph theory, the $b_{nm}$ factors may be
represented by edges (between nodes $n$ and $m$); then the clusters
are the sets of nuclei in each 
connected subgraph (each of these being involved in an
interdependent process).  Figure~\ref{FigConnectedCluster}(b) illustrates 
an example of this.  
In this way we begin to relate our concept of clusters of 
interdependent processes to Eq.~(\ref{v_E}).

\begin{figure}
\includegraphics[width=3.4in]{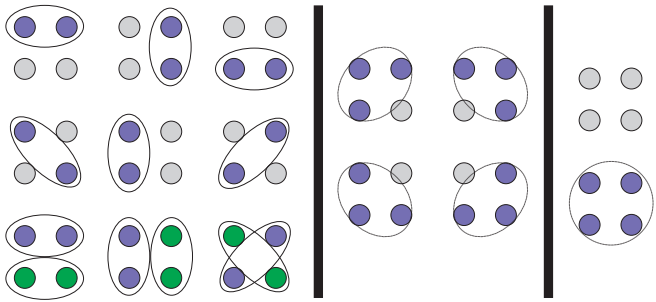}
\caption{
\label{figClusterDecomposition}
Set of all possible sets, $\{C_i\}$,  
of disjoint contributing clusters contained
in a set, ${\cal S}$, of four nuclei as an example.  
Contributing clusters are of
size $2$ or greater (a single nucleus gives no contribution on its
own).
The cases on the left involve 2-nuclei, middle ones involve 3-nuclei,
and the ones on the right are the trivial cases of $\{C_i\}=\emptyset$
or $\{C_i\} = \{{\cal S}\}$.
Such possibilities are iterated over in the summation of
Eq.~(\ref{vS_decomposition}).  Equation~(\ref{vS_decomposition_explicit1})
removes the $\{C_i\}=\emptyset$ case and
Eq.~(\ref{vC'_recursive}) removes
the $\{C_i\}=\{{\cal S}\} = \{{\cal C}\}$ case from their
respective summations.
}
\end{figure}

Each term in such a sum of products expansion will involve some
disjoint set of clusters.  Let us assume that processes involving
disjoint clusters are truly independent.  For a 
particular cluster, ${\cal C}$, we should then be able to
factor out a unique cluster contribution from those terms which
involve cluster ${\cal C}$.  This is the basis for a decomposition of
$v_E(\tau)$ into cluster contributions.
It will be useful to generalize this by defining 
$v_{\cal S}(\tau)$ to be the solution 
to the Hahn echo, $v_E(\tau)$ [Eq.~(\ref{v_E})], 
when only including the nuclei in some 
set ${\cal S}$.  We will use $v_{\cal C}'(\tau)$ to represent the
contribution from cluster ${\cal C}$.  Appendix
\ref{appendix_clusterIndependence} proves that the cluster
contributions are independent of each other in a way that allows for
the following decomposition of $v_{\cal S}(\tau)$:
\begin{eqnarray}
\label{vS_decomposition}
v_{\cal S}(\tau) &=& \sum_{
\substack{
\left\{{\cal C}_i\right\}~\mbox{\scriptsize{disjoint}}, \\
{\cal C}_i~\ne~\emptyset,~{\cal C}_i~\subseteq~{\cal S}
}}
\prod_i v_{{\cal C}_i}'(\tau) \\
\label{vS_decomposition_explicit1}
&=& 1 + \sum_{
\substack{
\left\{{\cal C}_i\right\} \ne \emptyset ~\mbox{\scriptsize{disjoint}},
\\
{\cal C}_i~\ne~\emptyset,~{\cal C}_i~\subseteq~{\cal S}
}}
\prod_i v_{{\cal C}_i}'(\tau),
\end{eqnarray}
where the summation of Eq.~(\ref{vS_decomposition}) is over all possible 
sets, $\{{\cal C}_i\}$, 
of disjoint nonempty
clusters, ${\cal C}_i$, each of which is contained in or equal 
to ${\cal S}$.  In other words, it iterates over all possible ways of
dividing any part of ${\cal S}$ into disjoint clusters as 
depicted in Fig.~\ref{figClusterDecomposition}.  The product is over all
clusters in each set.  Despite the index, $i$, the order is irrelevent
and permutations do not count as distinct cases.
Extracting the trivial $\{{\cal C}_i\} = \emptyset$ term yields 
Eq.~(\ref{vS_decomposition_explicit1}), shown explicitly 
to avoid confusion or ambiguity.
The unique existence of such a decomposition follows from the fact
that any $v_{\cal C}'(\tau)$ must be well-defined 
independent of any nuclei outside of ${\cal C}$ which is
proven in Appendix \ref{appendix_clusterIndependence}.

We can use
Eq.~(\ref{vS_decomposition}) itself to obtain an unambiguous expression
for any $v_{\cal C}'(\tau)$.  We do this by applying
Eq.~(\ref{vS_decomposition}) to the case in which ${\cal S} = {\cal
  C}$ and pulling out the term, from the summation, 
in which $\{{\cal C}_i\} = \{{\cal C}\}$ leaving only sets
in which all ${\cal C}_i~\ne~{\cal C}$:
\begin{equation}
v_{\cal C}(\tau) = v_{\cal C}'(\tau) +
\sum_{
\substack{
\left\{{\cal C}_i\right\}~\mbox{\scriptsize{disjoint}}, \\
{\cal C}_i \ne \emptyset,~{\cal C}_i~\subset~{\cal C}
}}
\prod_i v_{{\cal C}_i}'(\tau),
\end{equation}
so that
\begin{equation}
\label{vC'_recursive}
v_{\cal C}'(\tau) = v_{\cal C}(\tau) -
\sum_{
\substack{
\left\{{\cal C}_i\right\}~\mbox{\scriptsize{disjoint}}, \\
{\cal C}_i \ne \emptyset,~{\cal C}_i~\subset~{\cal C}
}}
\prod_i v_{{\cal C}_i}'(\tau).
\end{equation}
Equation~(\ref{vC'_recursive}) provides a recursive
definition of a cluster contribution.  Starting with the computation
of $v_{\cal C}(\tau)$, which is feasible to calculate directly for
small clusters, one must subtract terms that involve multiple
independent processes and processes that do not involve all of the
nuclei in ${\cal C}$.  It is a direct consequence of the
decomposition given by Eq.~(\ref{vS_decomposition}).

To ensure that Eq.~(\ref{vC'_recursive}) is well-understood,
we show more explicit results for clusters of size one through four.  
Because a single isolated nucleus cannot contribute to 
spectral diffusion,
$v_{{\cal C}_1}'(\tau) = v_{{\cal C}_1}(\tau) - 1 = 0$ when 
$|{\cal C}_1| = 1$.  
It follows that for $2$-clusters, 
$v_{{\cal C}_2}'(\tau) = v_{{\cal C}_2}(\tau) - 1$ 
(with $|{\cal C}_2| = 2$), having no contributing
proper subclusters.  For
$3$-clusters, we must subtract off contributions from contained pairs:
\begin{equation}
v_{{\cal C}_3}'(\tau) = v_{{\cal C}_3}(\tau) - 1 - 
\sum_{
\substack{
{\cal C}_2~\subset~{\cal C}_3,\\
|{\cal C}_2| = 2}}
v_{{\cal C}_2}'(\tau).
\end{equation}
For $4$-clusters, we must also subtract off contributions from
contained $3$-clusters and the products of contributions from contained
disjoint pairs:
\begin{eqnarray}
\nonumber
v_{{\cal C}_4}'(\tau) &=& v_{{\cal C}_4}(\tau) - 1 - 
\sum_{
\substack{
{\cal C}_2~\subset~{\cal C}_4,\\
|{\cal C}_2| = 2}} v_{{\cal C}_2}'(\tau) -
\sum_{\substack{
{\cal C}_3~\subset~{\cal C}_4,\\
|{\cal C}_3| = 3}} v_{{\cal C}_3}'(\tau) \\
\label{vC'_4cluster}
&&~{}-
\frac{1}{2} 
\sum_{
\substack{
{\cal A}~\bigcup~{\cal B}~=~{\cal C}_4,\\
|{\cal A}| = |{\cal B}| = 2}}
v_{\cal A}'(\tau) v_{\cal B}'(\tau).
\end{eqnarray}
The factor of one-half in the last term is needed to compensate for
the fact that ${\cal A}$ and ${\cal B}$ may be swapped in the
summation; it is only a consequence of the notation used here (where
${\cal A}$ and ${\cal B}$ are interchangeable labels).

\subsubsection{Ideal cluster expansion}
\label{IdealClusterExpansion}
We are now able to compute cluster contributions to be used in
the evaluation of our cluster expansion.
Revising Eq.~(\ref{vS_decomposition}) slightly, we may write the following
expression for the ideal cluster expansion up to $k^{\mbox{\scriptsize th}}$
order:
\begin{equation}
\label{vE_cluster_ideal}
v_{E}^{(k)}(\tau) =
\sum_{
\substack{
\left\{{\cal C}_i\right\}~\mbox{\scriptsize{disjoint}}, \\
{\cal C}_i \ne \emptyset,~\lvert{\cal C}_i\rvert \leq k
}}
\prod_i v_{{\cal C}_i}'(\tau).
\end{equation}
In order to estimate the error of the $k^{\mbox{\scriptsize th}}$ order of the
expansion, we can compare it with the $(k+1)^{\mbox{\scriptsize th}}$ order
which must include contributions from $k+1$ sized clusters.
One way to convert $v_{E}^{(k)}(\tau)$ into $v_{E}^{(k+1)}(\tau)$
is to add additional terms to the sum in which
we replace any $k$-cluster contribution of an existing term with any
$(k+1)$-cluster contribution generated by adding one neighboring nucleus
to the original $k$-cluster.  In doing so, a replacement must be made because the original
$k$-cluster becomes disqualified when we introduce the new 
$(k+1)$-cluster which contains it (due to the requirement that the clusters be
disjoint).  
This approach will account for all new sets of $\{{\cal C}_i\}$
containing a $(k+1)$-cluster (since any $(k+1)$-cluster can be made by
adding a nucleus to a $k$-cluster); however, cases will be overcounted
because many $k$-clusters can be used to build the same $(k+1)$-cluster.
This is unimportant because our goal now is to estimate
the error of $v_{E}^{(k)}(\tau)$ relative to
$v_{E}^{(k+1)}(\tau)$ and overestimating
this error is just as good.  Proceeding along these lines, we first
separate out the $k$-cluster contributions:
\begin{equation}
v_{E}^{(k)}(\tau) =
\sum_{
\substack{
\left\{{\cal C}_i, {\cal D}_j\right\}~\mbox{\scriptsize{disjoint}}, \\
0 < \lvert{\cal C}_i\rvert < k,~\lvert{\cal D}_j\rvert = k
}}
\prod_{i} v_{{\cal C}_i}'(\tau) \prod_{j} v_{{\cal D}_j}'(\tau).
\end{equation}
This performs the same summation over sets of disjoint clusters
as in Eq.~(\ref{vE_cluster_ideal}) except that we label 
$k$-clusters as ${\cal D}_j$
and the smaller clusters as ${\cal C}_i$.
With these $k$-clusters now set apart, we can
estimate the error of $v_{E}^{(k)}(\tau)$ relative to
$v_{E}^{(k+1)}(\tau)$ by noting that the sum of all $(k+1)$-cluster
contribution replacements of $v_{{\cal D}_j}'(\tau)$ are roughly
${\cal O}(\lambda L) \times \lvert v_{{\cal D}_j}'(\tau) \rvert$.
Recall that $\lambda$ was introduced as a perturbation parameter such
that a cluster contribution of size $k$ scales as ${\cal
  O}(\lambda^k)$, and $L$ is the average number of neighbors so
that there are, roughly speaking, ${\cal O}(L)$ $(k+1)$-clusters that
may be built out of one $k$-cluster.  Thus
\begin{eqnarray}
\nonumber
v_{E}^{(k+1)}(\tau) &=&
\sum_{
\substack{
\left\{{\cal C}_i, {\cal D}_j\right\}~\mbox{\scriptsize{disjoint}}, \\
0 < \lvert{\cal C}_i\rvert < k,~\lvert{\cal D}_j\rvert = k
}}
\prod_{i} v_{{\cal C}_i}'(\tau) \times \\
&& \prod_{j} v_{{\cal D}_j}'(\tau) \left[1 + {\cal O}(\lambda L)\right].
\end{eqnarray}
If we explicity include these $(k+1)$-clusters, they would have
relative corrections of ${\cal O}(\lambda L)$ to account for 
$(k+2)$-clusters and so forth.
This provides a more rigorous argument for our previous assertion that
the cluster expansion converges when $\lambda L \ll 1$.

\subsubsection{Practical implementation of the cluster expansion}
\label{PracticalClusterExpansion}
\begin{figure}
\includegraphics[width=3.4in]{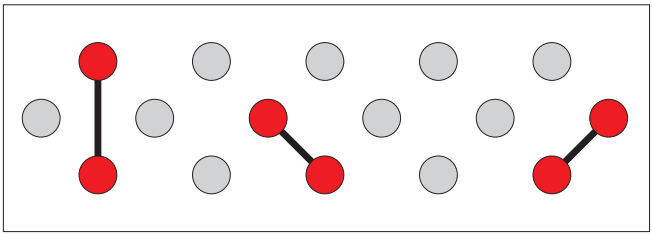}
\caption{
\label{schematic_pairs}
One possible combination of simultaneously
included pair contributions.  The red/dark circles are the
nuclei whose processes are being considered.
}
\end{figure}

Equation (\ref{vE_cluster_ideal}) directly implements the conceptual 
cluster expansion as described in Sec.~\ref{ConceptualClusterExpansion}
(the inclusion of contributions from all clusters up to size $k$); 
however, it is impractical for calculating results in large systems.
At the lowest nontrivial order, we would need to sum over all possible
products of disjoint pair contributions; for example, 
Fig.~\ref{schematic_pairs} 
depicts one such combination of disjoint pairs.
The number of such possibilities likely grows exponentially with the
problem size (we have not bothered to prove this rigorously 
or look it up, but certainly the scaling with problem size is horrendous).
However, we can effectively obtain all possible
combinations by making products of the form 
$\prod_{{\cal C}} \left[1+v_{\cal C}'(\tau)\right]$.
Distributing through a given factor yields the possibility of
excluding, via the $1$ term, or including, via
the $v_{\cal C}'(\tau)$ term, that cluster.  
Therefore such a product gives the sum of all
possible combinations of simultaneous cluster processes (for the
clusters included in the product).  Unfortunately, this will yield
combinations that involve overlapping clusters (that are therefore
not independent).  These overlapping clusters will introduce an error
that, in principle, may be corrected in successive orders of an approximation.

\begin{figure}
\includegraphics[width=3in]{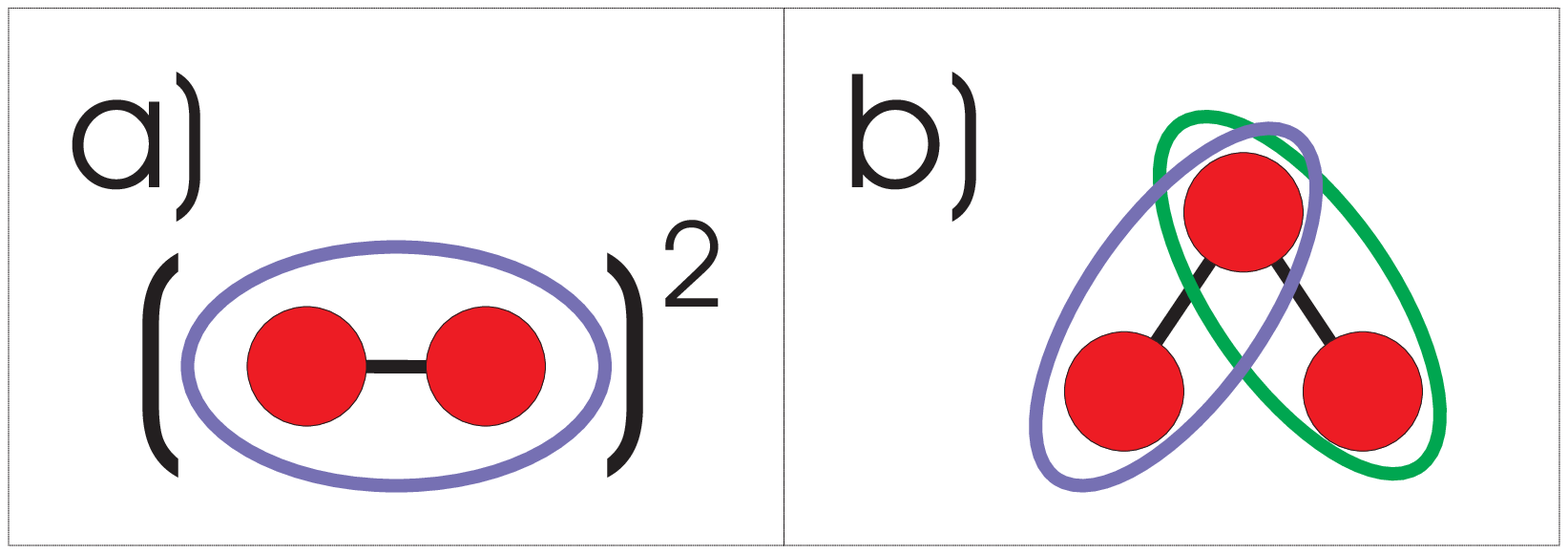}
\caption{
The practical implementation of the cluster expansion approximates the
ideal cluster expansion up to errors resulting from overlapping
clusters.  At the lowest order, such errors involve overlapping
pairs.  (a) A single pair multiplied by itself (i.e., a
pair overlapping itself) and (b) two pairs overlapping by sharing a nucleus.
\label{figOverlappingPairs}
}
\end{figure}

With this approach, the lowest nontrivial order of the expansion may
be implemented with
\begin{equation}
\label{vE2_prod_approx}
v_E^{(2)}(\tau) \approx \prod_{|{\cal C}_2| = 2} \left[1 + v_{{\cal C}_2}'(\tau)\right],
\end{equation}
producing all combinations of pair contributions along with some extraneous
terms, such as overlapping pairs as depicted 
in Fig.~\ref{figOverlappingPairs}(b).
For a moment, let us disregard these erroneous terms and consider the
consequence of this approximation.  If we take the logarithm of both sides,
we can convert the product on the right-hand side of
Eq.~(\ref{vE2_prod_approx}) into a convenient sum:
\begin{eqnarray}
\ln{\left(v_E^{(2)}(\tau)\right)} &\approx&
\sum_{|{\cal C}_2| = 2} \ln{\left(1 + v_{{\cal C}_2}'(\tau)\right)} \\
\label{vE2_TaylorLn}
&\approx& \sum_{|{\cal C}_2| = 2} v_{{\cal C}_2}'(\tau) 
\left[1 + {\cal O}\left(v_{{\cal C}_2}'(\tau)\right)\right],
\end{eqnarray}
where Eq.~(\ref{vE2_TaylorLn}) follows from the Taylor expansion of
$\ln{\left(1 + v_{{\cal C}_2}'(\tau)\right)}$ for 
$\lvert v_{{\cal C}_2}'(\tau) \rvert~\ll~1$ which we will shortly
justify in a self-consistent way.  If we assume that
$\lvert v_{{\cal C}_2}'(\tau) \rvert$ is small for all (or most) of the
${\cal C}_2$ pairs, then
\begin{eqnarray}
\label{vE2_expFormApprox}
v_E^{(2)}(\tau) &\approx& \exp{\left[\Sigma_2(\tau)\right]}, \\
\Sigma_2(\tau) &=& \sum_{|{\cal C}_2| = 2} v_{{\cal C}_2}'(\tau).
\end{eqnarray}
It is easy to show, from Eq.~(\ref{v_E}), that $v_{\cal S}(\tau)$ is always
real and that $-1 \leq v_{{\cal C}_2}(\tau) \leq 1$; therefore,
 $-2 \leq \left[v_{{\cal C}_2}'(\tau) = v_{{\cal C}_2}(\tau)
 - 1\right] \leq 0$.  Because 
$v_{{\cal C}_2}'(\tau=0) = v_{{\cal C}_2}(\tau=0) - 1 = 0$, 
$\Sigma_2(\tau = 0) = 0$ and becomes increasingly negative (initially
at the very least) as $\tau$ is increased.  For a large system, we
expect $\Sigma_2(\tau)$ to decrease monotonically to a value that is
${\cal O}(-N)$ [i.e. $v_{{\cal C}_2}'(\tau)$ have become random] so that
Eq.~(\ref{vE2_expFormApprox}) exhibits a decay form.  
The interesting part of the decay occurs when $\Sigma_2(\tau) \gtrsim
-1$ so that $v_E^{(2)}(\tau) \gtrsim e^{-1}$.  When this is true,
the average pair contribution will be, at most, ${\cal O}(1/N)$,
self-consistently justifying the approximation of
Eq.~(\ref{vE2_expFormApprox}) relative to
Eq.~(\ref{vE2_TaylorLn}) when $N$ is large (as it is for systems 
of interest).
Increasing $\tau$ much beyond this point will bring us to the tail of
the decay in which $v_E(\tau) \approx v_E^{(2)}(\tau) \ll 1$.  To
state this in a physically intuitive way, the decoherence of spectral
diffusion is caused by many nuclei collectively such that each potentially 
flip-flopping nuclear pair contributes only a small amount to the
overall dephasing before coherence is completely lost. 

For practical purposes [when $v_E(\tau)$ is not terribly
small], we thus regard each cluster contribution to be
${\cal O}(1/N)$.
Now let us discuss the extraneous overlapping pairs of 
Eq.~(\ref{vE2_prod_approx}) that we have thus far disregarded.  We can
now think of these cases, and their corrections, in orders of $1/N$
which increase with the number of overlapping clusters.  The lowest order
correction will therefore remove cases of two pairs that overlap with
each other.  For any given pair, there are ${\cal O}(L)$ pairs that
can overlap with it, each of which has a contribution of 
${\cal O}(1/N)$ as discussed above.  Therefore in estimating the
error of Eq.~(\ref{vE2_prod_approx}) we may write
\begin{equation}
\label{vE2_prod}
v_E^{(2)}(\tau) = \prod_{|{\cal C}_2| = 2} \left(1 + v_{{\cal C}_2}'(\tau)
\left[1 + {\cal O}\left(\frac{L}{N}\right)\right]\right).
\end{equation}
Carrying this error over to Eq.~(\ref{vE2_expFormApprox}) and
including its inherent ${\cal O}(1/N)$ error, we have
\begin{equation}
\label{vE2_expForm}
v_E^{(2)}(\tau) = \exp{\left(\Sigma_2(\tau)
\left[1 + {\cal O}\left(\frac{1+L}{N}\right)\right]\right)}.
\end{equation}

Because a cluster contribution scales in orders of $\lambda$ as we
increase the cluster size, this approach may be used for higher order
cluster contributions provided that $\lambda \ll N$ (typically,
$\lambda \ll 1$ where the cluster expansion is applicable). 
Taking either $\lambda \ll 1$ or $\lambda \sim 1$,
we may write, as an extension of the above approach to higher orders,
\begin{eqnarray}
\label{vEk_lnForm}
\ln{\left(v_E^{(k)}(\tau)\right)} &=& \sum_{j=2}^{k} \Sigma_{j}(\tau)
\left[1 + {\cal O}\left(\frac{1+L}{N}\right)\right], \\
\label{Sigma_j}
\Sigma_j(\tau) &=& \sum_{|{\cal C}| = j} v_{{\cal C}}'(\tau).
\end{eqnarray}
Note that $\Sigma_k(\tau) \sim \Sigma_{k-1}(\tau) \times {\cal
  O}(\lambda L)$, since there are roughly ${\cal O}(L)$ times as many
  $k$-clusters as $(k-1)$-clusters and on average each $k$-cluster 
contribution, by the definition of $\lambda$, is ${\cal O}(\lambda)$
  times that of the average $(k-1)$-cluster.  With this in mind, we
  see that, under the cluster expansion,
  $\ln{\left(v_E(\tau)\right)}$ is effectively expanded, additively, in
  powers of $(\lambda L)$.

In addition to the expansion in cluster size, we may also successively
correct for the ${\cal O}(1/N)$ errors of overlapping clusters.
This is done by starting with the smallest number of overlapping
clusters of the smallest sizes; that is, start with the case of
two overlapping pairs (Fig.~\ref{figOverlappingPairs}).
Each additional cluster included in the set of overlapping
clusters being considered will multiply ${\cal O}(1/N)$ to the
correction, and each additional nucleus added to any cluster will
multiply ${\cal O}(\lambda L)$ to the correction.  For our purposes, we
will only consider the correction for two overlapping pairs as a
check to verify that the approximation made in Eq.~(\ref{vEk_lnForm})
is valid.

There are two cases to consider for this lowest order
correction of overlapping clusters: 
the same pair multiplied by itself [Fig.~\ref{figOverlappingPairs}(a)] 
which was introduced by the 
approximation of Eq.~(\ref{vE2_TaylorLn}), and
two different pairs that overlap [Fig.~\ref{figOverlappingPairs}(b)]
which originates from Eq.~(\ref{vE2_prod_approx}).
These cases are, respectively, elimated, to
lowest order (two pairs and only two pairs that overlap), 
by adding the following to Eq.~(\ref{vEk_lnForm}):
\begin{eqnarray}
\label{sigma2_star}
\Sigma_2^*(\tau) &=& -\frac{1}{2} \sum_{|{\cal C}_2| = 2} 
\left[v_{{\cal C}_2}'(\tau)\right]^2, \\
\label{sigma3_star}
\Sigma_3^*(\tau) &=& - \frac{1}{2}\sum_{\substack{
\left\lvert{\cal A} \bigcup {\cal B}\right\rvert = 3, \\
\left\lvert{\cal A}\right\rvert = \left|{\cal B}\right| = 2
}}v_{\cal A}'(\tau) v_{\cal B}'(\tau),~~~ 
\end{eqnarray}
so that
\begin{eqnarray}
\label{vE_lnForm_2pairCorrected}
\ln{\left(v_E^{(k)}(\tau)\right)} &=& 
\Sigma_{2}(\tau) + \Sigma_{2}^{*}(\tau) + \Sigma_{3}^{*}(\tau) \\
\nonumber
&&{} +
\sum_{j=3}^{k} \Sigma_{j}(\tau)
\left[1 + {\cal O}\left(\frac{1+L}{N}\right)\right].~~~
\end{eqnarray}
Exponentiating Eq.~(\ref{vE_lnForm_2pairCorrected}) then expanding
and distributing this exponential into a sum of products form
will yield the
sum of all products of disjoint cluster contributions, as in
Eq.~(\ref{vE_cluster_ideal}), plus extraneous
terms of overlapping clusters.  However, all cases of only two pairs 
overlapping with each other (including a pair multiplied by itself)
will be removed as a result adding in $\Sigma_{2}^{*}(\tau)$ and
$\Sigma_{3}^{*}(\tau)$.  There
will remain higher order errors with more than two overlapping
clusters or overlapping clusters larger than pairs; in fact, 
additional higher order errors are
introduced by the $\Sigma_{2}^{*}(\tau)$ and $\Sigma_{3}^{*}(\tau)$
corrections itself.  For this
reason, it is difficult to derive higher order corrections (you must
correct errors introduced by lower order corrections).  We can, however,
regard this lowest order correction as an estimate of the error caused by
these extraneous overlapping clusters:
\begin{eqnarray}
\label{vE_practicalCluster}
\ln{\left(v_E^{(k)}(\tau)\right)} &=& 
\sum_{j=2}^{k} \Sigma_{j}(\tau) + {\cal
  O}\left(\Sigma^{*}(\tau)\right), \\
\label{Sigma_star}
\Sigma^{*}(\tau) &=&\Sigma_{2}^{*}(\tau) + \Sigma_{3}^{*}(\tau).
\end{eqnarray}
Note that $\Sigma_{2}^{*}(\tau)$ and $\Sigma_{3}^{*}(\tau)$ are both
$\leq 0$ and therefore add constructively
(otherwise we would want to take absolute values in order to estimate
the error conservatively).
Fortunately,
calculations of $\Sigma^{*}(\tau)$ indicate that it is a minor
correction for practical purposes.  Such calculations verify the
argument that these are ${\cal O}(1/N)$ errors [at least for practical
values of $\tau$ for which $v_E(\tau) \gtrsim e^{-1}$].

\subsection{Cluster expansion in summary}
The cluster expansion method that we have developed in this section is
very powerful and very general.  The disjoint cluster decomposition
[Eq.~(\ref{vS_decomposition})] could be used to take the trace of any
evolution operators described by Hamiltonians with pairwise (or even
higher order) interactions.  This decomposition may then 
be used to form an expansion [Eq.~(\ref{vE_cluster_ideal})]
that converges when the sum of cluster contributions decreases with
cluster size (i.e., $\lambda L \ll 1$).
In order to practically compute this
expansion for a large system, we need to use 
approximations such as Eq.~(\ref{vE2_prod}) or Eq.~(\ref{vEk_lnForm}) which
have the additional requirement that each cluster
contribution be small, e.g., {\cal O}(1/N), 
so that extraneous overlapping clusters arising from these
approximations are small.
This is, in principle, a formally exact, systematic expansion and its
convergence may be tested by comparing
$\Sigma_j(\tau)$ for at least $j=2,3,~\mbox{and}~4$ as well as 
$\Sigma^{*}(\tau)$.  It is important to compute $\Sigma_4(\tau)$ as
well as $\Sigma_3(\tau)$
because a symmetry of the problem eliminates all odd orders of $\lambda$
as we will show for both perturbation theories in 
Sec.~\ref{perturbation_theories}; therefore, $3$ cluster contributions are
actually ${\cal O}(\lambda^4)$.

We conclude this section by remarking that, besides being elegant
and useful for understanding the expansion, the natural logarithm form 
of the Hahn echo given by Eq.~(\ref{vE_practicalCluster}) 
has the advantage that
it is convenient to compute $\Sigma_j(\tau)$ and $\Sigma^{*}(\tau)$
using Monte Carlo techniques.  Rather than computing the full sum,
randomly selected terms may be sampled and averaged in order to obtain an
estimate for each sum.  This can save a lot of computation time and
makes this method powerful for large, complicated systems.

\section{Dual Perturbation Theories}
\label{perturbation_theories}

The convergence of the cluster expansion, described in the previous
section, depends upon how cluster contributions scale with cluster
size.  In that section, we surmized that a perturbation theory 
could be invoked such that cluster contributions scale with 
cluster size in orders of some perturbation
parameter, $\lambda$;  that is, the contribution of a cluster of size
$k$ is ${\cal O}(\lambda^k)$.
In this way, the cluster expansion relies upon an underlying 
perturbation theory to justify or
explain its convergence.  
Alternatively, one may regard the cluster expansion as a means to
extend convergence of a perturbative expansion to large systems (i.e., even
when $N \gg \lambda$).
In any case, in order to form a connection between a perturbation
theory and the cluster expansion, the contribution of a
cluster of size $k$ must be shown to
have a minimum perturbative order of $\lambda^k$
(i.e., lower orders cancel out).
In this section, we present two such perturbation theories
which are applicable in complimentary regimes (and thus, ``dual''). 

\subsection{Two theories in complementary regimes}
\label{dual_perturb_regimes}
Although otherwise complementary, both perturbation 
theories require that $\tau \ll \Max{(b_{nm})}^{-1}$.  
However, in problems we have considered
$\tau_D \ll  \Max{(b_{nm})}^{-1}$ where $\tau_D$ is the decay time,
and therefore this constraint has no practical consequence.  
The first perturbation theory we will present is an expansion in
orders of $\tau$.  For this theory, we can effectively assign
 $\lambda = \Max{(b_{nm})} \tau$ as the perturbation parameter used in
 Sec.~\ref{cluster_method}.  It is apparent that
$\tau \ll  \Max{(b_{nm})}^{-1}$ for $\lambda$ to be small.
Besides this requirement, we will see that this perturbative expansion 
is generally convergent in the regime in which $|A_n - A_m| \ll b_{nm}$. 
The second perturbation theory, which we call the dipolar
perturbation technique, will treat ${\cal H}^B$ of Eq.~(\ref{H_Bnm}) as a
perturbation to the free evolution Hamiltonian.  
For this theory, $\lambda \sim b_{nm} / |A_n - A_m|$ and
is therefore convergent in the
opposite regime as the $\tau$-expansion with respect to $|A_n - A_m|$
versus $b_{nm}$; that is, $b_{nm} \ll |A_n - A_m|$.

For convenience, we define
\begin{equation}
c_{nm} = \frac{A_n - A_m}{4 b_{nm}},
\label{cnm}
\end{equation}
so that the $\tau$-expansion perturbation theory is said to be
applicable in the $c_{nm} \ll 1$ regime while the dipolar perturbation
is applicable in the $c_{nm} \gg 1$ regime.
The factor of $4$ in Eq.~(\ref{cnm}) really only makes sense in the
context of spin-$1/2$ nuclei, but it is irrelevant for the current
discussion.  Nuclei near the donor or center of the dot, where the
electron wave function is strongest so that hyperfine coupling dominates the
internuclear coupling [$\Max{(A_n)} / \Max{(b_{nm})} \sim
10^{3}$], can typically be classified in 
the $c_{nm} \gg 1$ regime.  Nuclei far from the donor or center
of the dot can typically be classified in the opposite regime.  A
consequence of our perturbative theories is that the extreme cases
will give negligible contributions (i.e., there is a ``frozen'' core
where nuclear flip-flops are suppressed by the strong hyperfine
coupling, and the electron's 
interactions with distant nuclei are too weak to be of
consequence).  In fact, it is arguable that clusters of the $c_{nm}
\sim 1$ regime give the strongest contributions which seems to
contradict our supposition that either of these perturbation theories
are applicable.  However, time plays a role that shifts the balance to
the $c_{nm} \gtrsim 1$ side enough to make the dipolar perturbation
theory particularly applicable in a way that causes the
cluster expansion to converge.  This is because the $c_{nm} > 1 $ regime
implies larger energy scales (with strong hyperfine coupling than the 
converse) which causes such
cluster contributions to operate on shorter time scales and dominate
the Hahn echo decay.  We would not be satisfied with the face value of
this argument, but we have performed a number of numerical
calculations that support this conclusion.  In particular, we have
compared the cluster expansion performed in two
ways: one in which cluster contributions are computed exactly, and the 
other in which these contributions are approximated by dipolar
perturbation theory in the lowest order.  
In order to give convergent results for the
latter calculation, a lower bound $c_{nm}$ cutoff must be imposed.
However, for a broad range of $c_{nm}$ cutoff values, the two
calculations agree very well.  This implies that dipolar perturbation
theory completely accounts for the Hahn echo decay.  The main importance of
the $\tau$-expansion perturbation theory is to justify the $c_{nm}$
cutoff by the argument that $c_{nm} \ll 1$ clusters are negligible
(although it will have other purposes as we discuss shortly).

We emphasize that the convergence of the
cluster expansion is ultimately proven by performing the calculations
and comparing the different orders of the expansion (as we do).
The dual perturbation theories serve to explain this convergence.

\subsection{Expansion in tau}
\label{tau_expansion}

Before we devised the cluster method of solving the spectral
diffusion problem, we attempted to obtain a solution, at short
times, by performing an expansion in $\tau$ directly.  This met with
failure by not converging (for the Si:P system
which we tested) even at very short times.  We now understand that the
reason for the failure of the direct $\tau$-expansion is a combination
of the large system size and the difference in the energy (time) scales of
the hyperfine coupling and dipolar coupling.
In the context of the cluster expansion, however, a perturbation in
orders of $\tau$ can be applicable.  The cluster method deals with
large systems correctly and can allow underlying perturbation theories to
exhibit themselves without being drowned out by a large $N \gg
\lambda^{-1}$.  

In Sec.~\ref{algebra}, we will describe the techniques that
we developed in order to analyze and compute this $\tau$-expansion.
These techniques are shown (in Appendices) as they may
be adapted to other problems for which they are more suitable.
In Sec.~\ref{tau_cluster}, we show that, in
the regime in which the expansion is convergent for small clusters,
this provides an underlying perturbation theory to justify the
convergence of the cluster expansion (i.e., providing the $\lambda$
used in Sec.~\ref{cluster_method}).  
Because of the difference in the energy (time) scales of
the hyperfine coupling and dipolar coupling,
this perturbation theory is not as relevant as the
dipolar perturbation described in Sec.~\ref{perturbation}; however,
we will see that consequences of the $\tau$-expansion are exhibited in
GaAs systems (Sec.~\ref{GaAsResults}).
Also,
Fig.~\ref{SiTauCluster} of section \ref{results}, 
provides independent verification of the cluster expansion by
using this expansion directly (not in the context of the cluster expansion) for
an artificial system in which it is applicable.

\subsubsection{Algebraic expansion}
\label{algebra}
An expansion in $\tau$ is performed by
 expanding the exponentials of $U_{\pm}(\tau)$ [Eq.~(\ref{Upm})] in
 Eq.~(\ref{v_E}) and collecting all distributed terms of the desired order in
$\tau$.  
One can additionally expand
 ${e}^{-{\cal H}^{n}/k_B T}$, from Eq.~(\ref{v_E_temp}), 
in powers of $T^{-1}$ but in the
 remaining discussion we will be taking the $T \rightarrow \infty$ limit.  
We developed a symbolic manipulation computer program that gives an
exact algebraic expression for any expansion order (although
computation time limits the expansions that are feasible).  The program
uses its own algebraic manipulation library created precisely for this
purpose.  Appendix \ref{appendix_ComputerAlgebra} 
describes the procedure used by the program to obtain this expression.

The computer program's results for $I=1/2$ were compared to
direct hand calculations (as a check) for up to
sixth order in $\tau$.  Actually, odd orders can be shown to be
zero since the result must be real and $\tau$ is always accompanied by
a factor of $i$.  The result is more compact when put in terms of
$c_{nm}$ as defined by Eq.~(\ref{cnm}).  In this form the result is
\begin{eqnarray}
\label{vE_tau}
v_{E}(\tau) & = & 1 - D_4 \tau^4 - D_6 \tau^6 + {\cal O}(\tau^8),\\
\label{tau4}
\label{vE_tau4}
D_4 & = &  4 c_{nm}^2 b_{nm}^4, \\
\label{vE_tau6}
-\frac{3}{4} D_6 &=& c_{nm}^4 b_{nm}^6 +c_{nm}^2 b_{nm}^6\\
\nonumber
&&+c_{nm}^2 b_{nm}^4 \left(b_{nk} - b_{mk}\right)^2 \\
\nonumber
&&+c_{nm} c_{nk} b_{nm}^2 b_{nk}^2
\left[b_{nk} \left(b_{nm} + b_{mk}\right) + b_{mk}^2\right],
\end{eqnarray}
where distinct summation over indices is implied for each term.  Written in this
form,
we can surmise that the $\tau$ expansion
gives convergence for $\Max{(b_{nm})} \tau \ll 1$ in the regime in
which most pairs satisfy $c_{nm} \ll 1$.  As indicated above,
typical spectral diffusion problems have many pairs that 
satisfy $c_{nm} \gg 1$ instead.

This program has computed algebraic expressions up to
${\cal O}(\tau^{10})$ for both $I=1/2$ and $I=3/2$.  In the $I=1/2$ case, 
there are $111$ eighth order terms, and $1200$ tenth order terms.  
Terms of order $\tau^8$ have up to four distinct nuclear index
labels; terms of order $\tau^{10}$ have up to five distinct nuclear
index labels.  The trend that the number of
distinct nuclear indices is half the order of $\tau$ will be used in
Sec.~\ref{tau_cluster} to relate the tau expansion to the cluster expansion.

After computing these algebraic expressions, there is an additional
challenge of efficiently computing the coefficients of $\tau$ from these
expressions for specific values of $A_n$
and $b_{nm}$ in order to obtain results such as the $\tau$-expansion
plots of Fig.~\ref{SiTauCluster}.  Appendix
\ref{appendix_tau_computation} describes how this is done.

\subsubsection{Relating the tau-expansion to clusters}
\label{tau_cluster}
The trend seen above is that the maximum number of distinct nuclear
index labels is half of the order of $\tau$.  This provides a
connection to the cluster expansion.  By definition, 
a cluster contribution must involve all of the nuclei in that
cluster. 
The number of distinct nuclear
indices for a cluster contribution is thus equal to the cluster size.
Therefore, clusters of size $2$ are, at a minimum, fourth order in $\tau$;
clusters of size $3$ are ${\cal O}(\tau^6)$, etc.  
Even if we cannot prove that this trend
continues beyond ${\cal O}(\tau^{10})$ we know that clusters of size greater or
equal to $5$ are at least ${\cal O}(\tau^{10})$.  This is a key formal
connection between the $\tau$-expansion and the cluster expansion.

We can therefore relate orders of $\lambda$ used in 
Sec.~\ref{cluster_method} to orders of $\tau$ (up to 5-clusters at least).
As long as the $\tau$ expansion is convergent for small $N$, the
 cluster expansion can extend its convergence for large $N$.
As discussed above, this convergence in $\tau$ occurs for 
$\Max{(b_{nm})} \tau \ll 1$ in the regime in which most pairs 
satisfy $c_{nm} \ll 1$.  We also observed, from the argument that
 $v_E(\tau)$ must be real, that no odd orders of $\tau$ are present.
 Therefore, as discussed in Sec.~\ref{cluster_method}, the next
 order of the cluster expansion beyond pairs should include
 3-clusters and 4-clusters.

\subsection{Dipolar perturbation expansion}
\label{perturbation}

An expansion in $\tau$ is not well-suited, on its own, to practical problems of
interest due to the many nuclear pairs satisfying $c_{nm} \gg 1$
[defined in Eq.~(\ref{cnm})].  
However, in the regime in which $c_{nm} \gg 1$, a
nondegenerate perturbation expansion in orders of $b_{nm}$ is
applicable.  We introduce a bookkeeping
parameter $\lambda$ (which we will later relate to the $\lambda$ used in
Sec.~\ref{cluster_method}, making a formal connection to the
cluster expansion)
 such that $\pm {\cal H}^{\pm}={\cal H}^0 \pm
\lambda {\cal H}'$.  Here the unperturbed Hamiltonian, ${\cal H}^0 =
\frac{1}{2}\sum_n A_n I_{nz}$, is diagonal in the nuclear spin $z$-basis, while ${\cal H}'=\frac{1}{\lambda}{\cal H}^B$ is the dipolar interaction rescaled
to have the same magnitude as ${\cal H}^0$.  
We note that $\lambda$ here is a purely formal artifice to
define a perturbation expansion using
standard perturbation theory in quantum mechanics (and later adapted
to the cluster expansion of Sec.~\ref{cluster_method} in order to
ensure convergence for large systems).  

\subsubsection{Perturbed Hamiltonian}
Using standard perturbation theory in quantum mechanics, we
 have the following recursive definitions
for the eigenvectors and eigenvalues of ${\cal H}^{\pm}$:
\begin{eqnarray}
\label{eigenvector_recursive}
\left|k_{\pm}\right> &=& \left|k^{0}\right> \pm \lambda \sum_{l \ne k}
\left|l^{0}\right> \times \\
\nonumber
&&
\frac{\left<l^{0}\right|{\cal H}'\left|k_{\pm}\right> - 
\left<l^{0}\right.\left|k_{\pm}\right>\left<k^{0}\right|{\cal H}'\left|k_{\pm}\right>}
{E_k^{(0)} - E_l^{(0)}}, \\
\label{eigenvalue_recursive}
E_k^{\pm} &=& E_k^{(0)} \pm \lambda \left<k^{0}\right|{\cal
  H}'\left|k_{\pm}\right>, \\
\label{E_0}
E_k^{(0)} &=& \frac{1}{2} \sum_n A_n \left<k^{0}\right| I_{nz} \left|k^{0}\right>
\end{eqnarray}
with the normalization convention that
\begin{equation}
\left<k^{0}\right|\left.k\right> =
\left<k^{0}\right|\left.k^{0}\right> = 1.
\end{equation}

Now we have
\begin{equation}
{\cal H}^{\pm} = \pm \sum_{k} E_k^{\pm}
\frac{\left|k_{\pm}\right> \left<k_{\pm}\right|}{\left<k_{\pm}\right.\left|k_{\pm}\right>},
\end{equation}
where the summation is over all possible states.  From Eq.~(\ref{Upm}),
\begin{equation}
U_{\pm}(\tau) = \sum_{k} 
\frac{\left|k_{\pm}\right>
  \left<k_{\pm}\right|}{\left<k_{\pm}\left.\right|k_{\pm}\right>} e^{\mp i E_k^{\pm} \tau}.
\end{equation}
Plugging this into Eq.~(\ref{v_E}) we have
\begin{eqnarray}
\label{perturb_amp}
v_E(\tau) &=& \sum_{i, j, k, l} \frac{C_{ijkl}}{M} \exp^{-\imaginary \omega_{ijkl} \tau}, \\
\label{Cijkl}
C_{ijkl} &=& \frac{F_{ijkl}}{D_{ijkl}}, \\
\label{perturb_numer}
F_{ijkl} &=&
  \left<l_-\right.\left|i_+\right>\left<i_+\right.\left|j_-\right>\left<j_-\right.\left|k_+\right>\left<k_+\right.\left|l_-\right>,
  \\
\label{perturb_denom}
D_{ijkl} &=&
\left<i_+\right.\left|i_+\right>\left<j_-\right.\left|j_-\right>\left<k_+\right.\left|k_+\right>\left<l_-\right.\left|l_-\right>,
\\
\label{omega}
\omega_{ijkl} &=& \omega_{ijkl}^{(0)} + \lambda \omega_{ijkl}', \\
\label{omega0}
\omega_{ijkl}^{(0)} &=& E_{i}^{0} - E_{j}^{0} - E_k^{0} + E_l^{0}, \\
\label{omega'}
\omega_{ijkl}' &=& \left<i^{0}\right|{\cal H}'\left|i_{+}\right> 
                 - \left<j^{0}\right|{\cal H}'\left|j_{-}\right>  \\
\nonumber
&&
                 - \left<k^{0}\right|{\cal H}'\left|k_{+}\right> 
                 + \left<l^{0}\right|{\cal H}'\left|l_{-}\right>.
\end{eqnarray}

We note that $C_{ijkl}$ and $\omega_{ijkl}'$ can be expanded in orders
of $\lambda \sim c_{nm}^{-1}$.

\subsubsection{Relating the dipolar perturbation expansion to clusters}
\label{perturb_cluster}
In order to show that the dipolar perturbation expansion can be used as
an underlying perturbation theory for the cluster expansion
(i.e., to supply the $\lambda$ of Sec.~\ref{cluster_method}), 
we need to show that for a given term in the $\lambda$
expansion of
Eq.~(\ref{perturb_amp}), there are at least as many orders of
$\lambda$ as there are nuclei involved (cluster size).  
In Appendix \ref{DipolarAndClusterSize}, 
we show this to be effectively the case 
in the $\Max{(b_{nm})} \tau \ll 1$ limit.  

One further point is that $v_E(\tau)$ is an even function of $\lambda$.
This is seen by noting that under the transformation $\lambda
\rightarrow -\lambda$ we have $U_{\pm} \leftrightarrow
U_{\mp}^{\dag}$ and therefore
\begin{eqnarray}
v_E(\tau) &=& \frac{1}{M}\left|\Tr{\left\{
U_{-}U_{+}U_{-}^{\dag}U_{+}^{\dag}
\right\}} \right| \\
&\rightarrow&  \frac{1}{M}\left|\Tr{\left\{
U_{+}^{\dag} U_{-}^{\dag} U_{+} U_{-}\right\}} \right| = v_E(\tau).
\end{eqnarray}
This is true because there is a symmetry between ``up'' and ``down'';
that is, $U_{-} \leftrightarrow U_{+}$ is symmetric
within the trace operation because 
$I_{nz} \rightarrow -I_{nz}~\forall~n$ is symmetric within the trace
operation.  A consideration of finite nuclear temperatures would
break this symmetry, but in our infinite nuclear temperature
approximation there will be no odd orders of $\lambda$.  Therefore, as
discussed in Sec.~\ref{cluster_method}, the next
 order of the cluster expansion beyond pairs should include
 3-clusters and 4-clusters.  
This was shown to be necessary in Sec.~\ref{tau_cluster}
when relating cluster sizes to orders of $\tau$, and it is also true
here when relating cluster sizes to orders of the dipolar perturbation expansion.

\section{Calculations and Results}
\label{results}

The information (i.e., the input) necessary to calculate the spectral diffusion Hahn
echo decay, given in Eq.~(\ref{v_E}) and approximated in the
cluster expansion by Eq.~(\ref{vE_practicalCluster}), are the $A_n$,
$b_{nm}$, and $I_n$ (nuclear spin magnitude) values.  The
internuclear dipolar coupling for $b_{nm}$ is given by Eq.~(\ref{bnm}).
For $A_n$ we use\cite{desousa03b}
\begin{eqnarray}
\label{An}
A_{n} &=&
\frac{8\pi}{3}\gamma_{S}\gamma_{I}\hbar|\Psi(\bm{R}_{n})|^{2} \\
\nonumber
&& - \gamma_{S}\gamma_{I}\hbar \frac{1 - 3
  \cos^{2}{\theta_n}}{\left|\bm{R}_n\right|}
\Theta(\left|\bm{R}_n\right| - r_0), 
\end{eqnarray}
The first term of Eq.~(\ref{An}) is the hyperfine coupling of
the $n^{\mbox{\scriptsize th}}$ nucleus to the electron, while the second term is its
dipolar coupling to the electron with $r_0$ being the effective range
of the wave function.  $\bm{R}_{n}$ is the position of the $n^{\mbox{\scriptsize th}}$
nucleus relative to the center of the electron wave function and
$\theta_{n}$ is its angle relative to the magnetic field direction.
While both of these terms were included in our 
calculations, the dipolar contribution of $A_n$ proves to be
negligible for the semiconductor spin quantum computer architectures.

\subsection{Phosphorus donor in silicon}
\label{Si:P}

Our first application is to consider the Hahn echo decay of the
electron spin of a phosphorus donor in natural silicon.\cite{chiba72,abe04,desousa03b}  Here $\Psi(\bm{R}_n)$
is the Kohn-Luttinger wave function of a phosphorus donor impurity in
silicon, as described in Ref.~\onlinecite{desousa03b}. Using this in
Eq.~(\ref{An}), we have\cite{desousa03b}
\begin{eqnarray} 
\label{An_Si}
A_{n} &=& \frac{16 \pi}{9}\gamma_{S}\gamma_{I}^{Si} \hbar
\eta \left[F_1(\bm{R}_n) \cos{(k_0 X_n)}\right. \\
\nonumber
&&\left.+F_3(\bm{R}_n) \cos{(k_0 Y_n)} + F_5(\bm{R}_n) \cos{(k_0
    Z_n)}\right]^2\\
\nonumber
&& - \gamma_{S}\gamma_{I}^{Si}\hbar \frac{1 - 3
  \cos^{2}{\theta_n}}{\left|\bm{R}_n\right|}
\Theta(\left|\bm{R}_n\right| - n a), \\
F_{1, 2}(\bm{r}) &=& \frac{\exp{\left[-\sqrt{\frac{x^2}{(n b)^2} +
          \frac{y^2 + z^2}{(n a)^2}}\right]}}{\sqrt{\pi (n a)^2 (n b)}},
\end{eqnarray}
with $\gamma_{S} = 1.76 \times 10^7 \mbox{(s G)$^{-1}$}$,
$\gamma_{I}^{Si} = 5.31 \times 10^3 \mbox{(s G)$^{-1}$}$, $n =
0.81$, $a = 25.09$~\AA, $b = 14.43$~\AA,
$\eta = 186$, $k_0 = (0.85) 2 \pi / a_{Si}$, and
$a_{Si} = 5.43$~\AA.  The Si nuclei are located on a diamond
lattice.\cite{kittel}  The central $^{31}$P nuclear spin does not contribute to spectral
diffusion because its hyperfine energy is significantly larger than
any of its neighbors, suppressing the spin flips by energy
conservation.   

In a natural sample of silicon, only a small fraction $f=4.67$\% 
of lattice sites have nonzero nuclear spin.  These are the
 spin-$1/2$ $^{29}$Si isotopes, therefore $I_n = 1/2$ 
 for all contributing nuclei.
We will use $\left\langle \Sigma_k(\tau) \right\rangle$ and
$\left\langle \Sigma_k^{*}(\tau) \right\rangle$ to denote
$\Sigma_k(\tau)$ and $\Sigma_{k}^{*}(\tau)$ averaged, respectively,
over isotopic configurations with a fraction, $f$, of $^{29}$Si.  We
will also use the convention that $\Sigma_k(\tau)$ and
$\Sigma_{k}^{*}(\tau)$ without these angle brackets gives the
$f=100\%$ result.  Thus
\begin{eqnarray}
\label{fScaled_Sigma_k}
\left\langle \Sigma_k(\tau) \right\rangle 
&=& f^k \Sigma_k(\tau), \\
\label{fScaled_Sigma_k_star}
\left\langle \Sigma_k^{*}(\tau) \right\rangle 
&=& f^k \Sigma_k^{*}(\tau),
\end{eqnarray}
where $\Sigma_k(\tau)$, $\Sigma_2^{*}(\tau)$, and $\Sigma_3^{*}(\tau)$
are given by Eqs.~(\ref{Sigma_j}), (\ref{sigma2_star}), and
(\ref{sigma3_star}), respectively, taking all nuclei to
be $^{29}$Si.  The fact that only a fraction, $f$, 
of these nuclei contribute to the diffusion is
accounted for by the $f^k$ factors in Eqs.~(\ref{fScaled_Sigma_k}) and (\ref{fScaled_Sigma_k_star})
because $f^k$ is the probability that all nuclei in a cluster of size $k$ have nonzero spin.

For the spin-$1/2$ nuclei that contribute to spectral diffusion in
natural silicon, we can write the following analytical solution for
pairs (2-clusters) using Eq.~(\ref{v_E}):
\begin{eqnarray}
\label{exact_pair_solution}
v_{nm}(\tau) &=& 1+ v_{nm}'(\tau)\nonumber\\
&=&1 - \frac{c_{nm}^2}{(1 + c_{nm}^2)^2}
\left[\cos{(\omega_{nm} \tau)} - 1\right]^2, \label{vnm}\\
\omega_{nm} &=& 2 b_{nm} \sqrt{1 + c_{nm}^2},
\end{eqnarray}
where $c_{nm}$ is given by Eq.~(\ref{cnm}).

\begin{figure}
\includegraphics[width=3in]{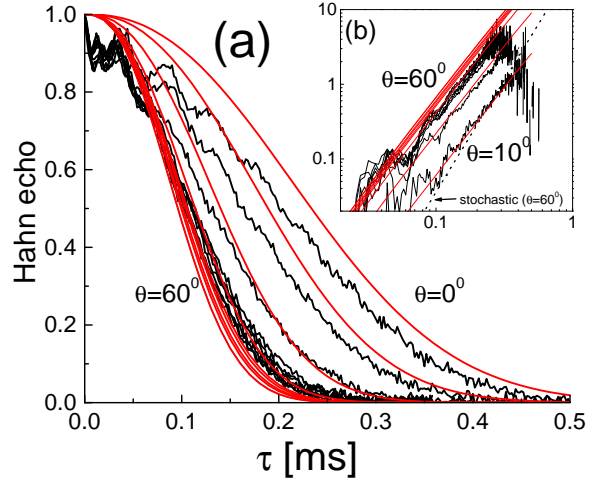}
\caption{
  Hahn echo decay $v_E(\tau,\theta)$ of a phosphorus donor electron
  spin in silicon due to the dipolar nuclear spin bath dynamics. 
(a) Theory (solid lines) and experiment (Ref.~\onlinecite{tyryshkinP}) is shown
  for several orientation angles of the magnetic field with respect to
  the crystal lattice, ranging from the [100] to the [110] direction
  ($\theta=0,10,20,\ldots,90$).  (b) Here we plot
  $-\ln{v_E}(\tau,\theta)+\ln{v_E(\tau,\theta=0)}$, allowing for the
  removal of any decoherence mechanism which is independent of
  $\theta$. The qualitative and quantitative agreement between theory
  and experiment is remarkable, in contrast to the stochastic approach
  (dashed).
\label{SiEcho}}
\end{figure}

Our numerical calculations of Hahn echo decay in the lowest order of
the cluster expansion, 
$v_E^{(2)}(\tau) = \exp{\left(\left\langle\Sigma_2(\tau)\right\rangle\right)}$
using Eqs.~(\ref{fScaled_Sigma_k}), (\ref{Sigma_j}), and (\ref{exact_pair_solution}),
are shown for several magnetic
field orientation angles in Fig.~\ref{SiEcho}(a)
with a direct
quantitative comparison to the experiment.\cite{tyryshkinP} 
The dipolar coupling [Eq.~(\ref{bnm})] contains an important anisotropy with respect to the
$\theta_{nm}$ angle formed between the applied magnetic field and the bond
vector linking the two spins ($\bm{R}_{nm}$). This property leads to a
strong dependence of spin echo decay when the sample is rotated with
respect to the applied B field direction.
The experimental data is taken for bulk natural silicon with phosphorus
doping concentration equal to $2\times 10^{15}$~cm$^{3}$
[\onlinecite{tyryshkinP}].  The high concentration of phosphorus donors leads
to an additional decoherence channel arising from the direct spin-spin
coupling between the electron spins that contribute to the echo.  This
contribution can be shown to add a multiplicative factor
$\exp{(-\tau/1\,\rm{ms})}$ to the Hahn echo.\cite{abragam62}
Because this contribution is independent of the orientation angle, we
can factor it out by subtracting the $\theta=0$ contribution from the
logarithm of the experimental data taken at angle $\theta$. The result
is shown in Fig.~\ref{SiEcho}(b) (log-log scale). Our theory seems to
explain the time dependence of the experimentally observed echo quite well.  
  This result
is to be compared with the recent stochastic theory of Ref.~\onlinecite{desousa03b} [Dashed line in Fig.~\ref{SiEcho}(b) shows the
stochastic calculation for $\theta=60^{\circ}$]. Although the
stochastic theory yields roughly correct coherence times in
order of magnitude, it fails qualitatively in explaining the time dependence 
[that is, the shape of the decay as can be seen from the incorrect
slope of the stochastic calculation in the log-log plot of
Fig.~\ref{SiEcho}(b)].  The
present method is able to incorporate all these features within a
fully microscopic framework, obtaining both qualitative and
quantitative agreement with experiment.

An important issue in the context of quantum information processing is the
behavior of spin coherence at the shortest time scales. The
experimental data\cite{tyryshkinP} in Fig.~\ref{SiEcho} reveals
several oscillatory features which are not explained by our current
method. These are echo
modulations arising from the anisotropic hyperfine coupling omitted in
Eq.~(\ref{H_A}).\cite{abe04}  This effect can be substantially reduced
by going to higher magnetic fields.  (In a quantum computer
$B\sim9$~Tesla will probably be required in order to avoid loss of
fidelity due to echo modulation.\cite{saikin03})  On the other hand,
spectral diffusion itself is essentially independent of magnetic field even
to extremely high magnetic field values ($B\sim 10$~Tesla).  
The echo modulation effect is 
expected to be absent in III-V materials,\cite{yablonovitch03} hence
our theory allows the study of spin coherence at time scales of great
importance for quantum information purposes, i.e., very short time
scales, as long as echo modulation effects are quantitatively
unimportant.

\begin{figure}
\includegraphics[width=3in]{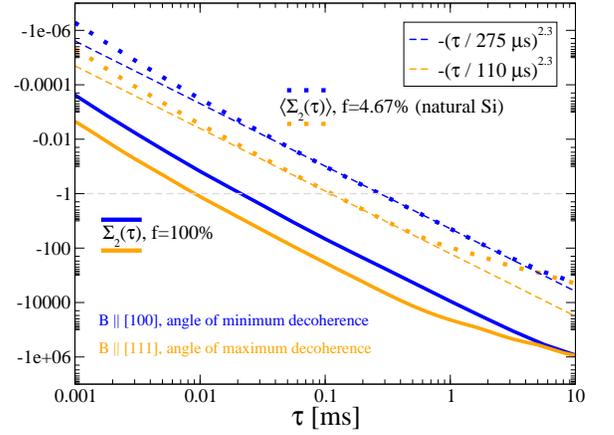}
\caption{
Lowest order results for the natural log of the Hahn echo,
$\ln{(v_E(\tau))} \approx \left\langle \Sigma_2(\tau) \right\rangle 
\propto f^2$,
for Si:P in a log-log plot.  
The solid lines give $\Sigma_2(\tau)$ with $f=1$.
Dotted lines give $\left\langle \Sigma_2(\tau) \right\rangle$ 
for natural Si ($f = 4.67\%$).  
In this log-log plot,
multiplying by $f^2$ simply shifts the curves vertically.
Isotopic purifucation would shift these curves up further.
The two magnetic field angles shown give extremal results.
Corresponding to $\theta$ angles in Fig.~\ref{SiEcho}, $B~||~[100]$ is
$\theta = 0^{\circ}$ and $B~||~[111]$ is $\theta \approx 54.7^{\circ}$.
Dashed lines fit the natural Si curves near their $-1$ values (where $v_E \sim
1/e$) with $\tau^{2.3}$ power law curves (linear in the log-log plot).
\label{logSi100AndNatSiFig}}
\end{figure}

Isotopic purification can reduce the value of $f$ (fraction of
$^{29}$Si nuclei).  Figure~\ref{logSi100AndNatSiFig} contains information
that is useful for understanding how the Hahn echo curves change as $f$ is
changed (i.e., lowered via isotopic purification).  In a log-log plot,
$\ln{(v_E(\tau))} \approx \left\langle \Sigma_2(\tau) \right\rangle \propto f^2$ 
simply shifts vertically when $f$ is
changed.  Figure~\ref{logSi100AndNatSiFig} shows both the $f$-independent
$\Sigma_2(\tau)$ (i.e., $f = 100\%$), 
and $\left\langle \Sigma_2(\tau) \right \rangle$ for natural Si ($f\sim
5\%$).  Results are
shown for magnetic field angles that yield the extremal slowest and
fastest decoherence.  For natural Si, in a wide range of $\tau$ about
$\tau_{1/e}$, where $v_E(\tau_{1/e}) = 1/e$, $\left\langle
\Sigma_2(\tau) \right\rangle$
matches $\tau^{2.3}$ curves very well.  In this range of $\tau$,
therefore, we may write
\begin{eqnarray}
\label{SiExpPower2.3}
v_E(\tau) &\approx& \exp{\left(-f^2 (\tau/\tau_0)^{2.3}\right)} \\
&=& \exp{\left(-(\tau/\tau_{1/e})^{2.3}\right)},
\end{eqnarray}
where
\begin{equation}
\label{time_f_scaling}
\tau_{1/e} = \tau_0 / f^{2/2.3} \propto f^{-0.87},
\end{equation}
providing a formula that allows us to adjust our Hahn echo curves to
other values of $f$ for a range of $\tau$ in which 
Eq.~(\ref{SiExpPower2.3}) is applicable.
Shortly after the original submission of this manuscript, experimental
data from Abe {\it et al.} appeared in the literature\cite{abe05} which shows
echo time scaling ranging between $f^{-0.86}$ and $f^{-0.89}$, in
remarkable agreement with our prediction [Eq.~(\ref{time_f_scaling})].  
Tyryshkin {\it et al.}\cite{tyryshkinP} report Si:P Hahn echo
decay forms of $\exp{(\tau^{2.4\pm0.1})}$, in agreement with
Eq.~(\ref{SiExpPower2.3}), with exception to magnetic field
orientations near the $[100]$ direction.  In the $[100]$ direction,
they report a form of $\exp{(\tau^{3.0\pm0.2})}$; the reason for this discrepancy 
remains unknown.

\begin{figure}
\includegraphics[width=3in]{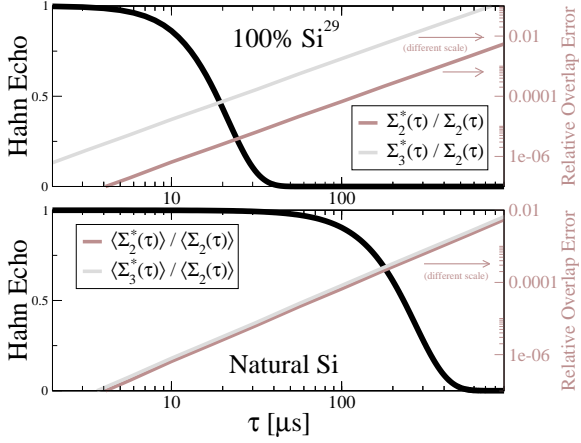}
\caption{
Relative errors (with scales on the right) 
to the log of the Hahn echo due to overlapping pairs
for both $100\%$ $^{29}$Si (top graph) and natural Si (bottom graph).
In these examples, $B~||~[100]$.
The Hahn echoes themselves are shown, as well, with the $0$ to $1$ scales on
the left.  All curves share the same logarithmic time ($\tau$) scale.
It is apparent that these relative corrections are very
small up to the tail of their respective echo decays.
\label{SiRelativeOverlapErrors}}
\end{figure}

We now check the convergence of our cluster expansion for this Si:P
system.  Using Eq.~(\ref{vE_practicalCluster}) and averaging over the
isotopic configurations yields
\begin{equation}
\label{vE_practicalSi}
\ln{\left(v_E^{(k)}(\tau)\right)} = 
\sum_{j=2}^{k} \left\langle \Sigma_{j}(\tau) \right\rangle + {\cal
  O}\left(\left\langle \Sigma^{*}(\tau) \right\rangle
 \right).
\end{equation}
This approximates the ideal cluster expansion [see Secs. 
\ref{IdealClusterExpansion} and \ref{PracticalClusterExpansion}] 
with an error that we may
estimate as $\left\langle \Sigma^{*}(\tau) \right\rangle =
\left\langle \Sigma_{2}^{*}(\tau) \right\rangle + 
\left\langle \Sigma_{3}^{*}(\tau)  \right\rangle$.  This error is
estimated by the correction needed to compensate for overlapping pairs
[either the same pair overlapping itself, $\Sigma_{2}^{*}(\tau)$, 
or two different pairs overlapping, $\Sigma_{3}^{*}(\tau)$]
in the approximation.  Figure~\ref{SiRelativeOverlapErrors} shows these
relative corrections, $\Sigma_{2}^{*}(\tau) / \Sigma_{2}(\tau)$ and 
$\Sigma_{3}^{*}(\tau) / \Sigma_{2}(\tau)$, to
$\ln{\left(v_E(\tau)\right)}$ for both $f=100\%$ and natural Si
($f \sim 5\%$).  The graphs also show the respective Hahn echoes (in a
log time scale which alters their appearance) to show that these
relative corrections are very small up to the tail of their respective
echo decays.  The argument, given in Sec.~\ref{PracticalClusterExpansion}, 
that $\Sigma^{*}(\tau)$ would be small was only applicable for
$v_E(\tau) \gtrsim e^{-1}$ so it is expected that this
approximation approaches failure out in the tail of the decay.  This
is irrelevant for practical purposes.

The expansion of Eq.~(\ref{vE_practicalSi}) is convergent where
$\left\langle \Sigma_{k+1}(\tau) \right\rangle \ll 
\left\langle \Sigma_{k}(\tau) \right\rangle$
(implying that $\lambda L \ll 1$ effectively).  
The $\Sigma_k(\tau)$ functions have been calculated (up to $k=5$)
using Monte Carlo
techniques with cluster contributions, $v_{\cal C}'(\tau)$, for
clusters that are larger than pairs,
calculated by numerically diagonalizing ${\cal H}^{\pm}$
[Eq.~(\ref{Hpm})].  For each $\Sigma_{k}(\tau)$ independently, the 
maximum distance between neighbors
and the maximum distance of nuclei to the donor is
increased for various Monte Carlo runs until convergence within a
desired precision is reached.  To speed up each Monte Carlo run, clusters
are chosen with a heuristic bias for those that have strong coupling 
between the constituent nuclei as well as a bias for clusters closer to
the donor.  
Appropriate weighting factors are used to counteract
these biases.

\begin{figure}
\includegraphics[width=3in]{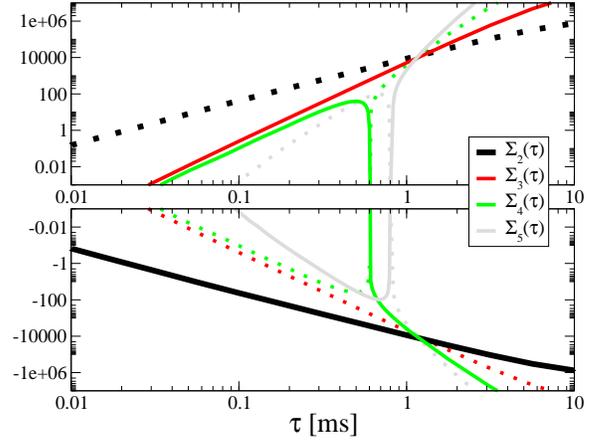}
\caption{
Successive orders of the cluster expansion for the natural log of the
Hahn echo [Eq.~(\ref{vE_practicalSi})], computed for Si:P and
$B~||~[100]$, for the $100\%$ $^{29}$Si theoretical scenario.  
The thick black line gives the
lowest order result, $\Sigma_2(\tau)$, and other solid lines give higher order
$\Sigma_k(\tau)$ results.  The dotted lines give the negative of
their corresponding functions 
provided to assist in the absolute value comparison of these higher
order corrections.  A failure of convergence occurs near $\tau \sim
1~\mbox{ms}$ where all of the curves are the same order of magnitude.
This occurs well into the tail of the decay, however, and therefore
has no practical consequence.
\label{logSiHigherOrder}}
\end{figure}

Figure~\ref{logSiHigherOrder} compares
$f$-independent (i.e., $f=100\%$) 
$\Sigma_{k}(\tau)$
functions in a dual (showing positive and negative values) log-log plot 
for Si:P with $B~||~[100]$.  In other words, it compares successive
orders of the expansion for the natural log of the Hahn echo,
$\ln{\left(v_E(\tau)\right)}$, with the $f$ dependence removed.
As one might anticipate by the fact that
$\Sigma_3(\tau)$ and $\Sigma_4(\tau)$ are both ${\cal O}(\lambda^4)$
(Sec.~\ref{perturbation_theories} proved that there are no odd orders of
$\lambda$ for either perturbation theory), they are similar orders of magnitude, at least for the
$0.03~\mbox{ms} < \tau < 1~\mbox{ms}$ range.  
Near $\tau \sim 1~\mbox{ms}$, however, the perturbation theory fails
[having the condition that $\Max{(b_{nm})} \tau \ll 1$] as we see 
that $\lvert\Sigma_4(\tau)\rvert$ surpasses
$\lvert\Sigma_3(\tau)\rvert$.
Interestingly, all orders approach the same order of magnitude near
$\tau \sim 1~\mbox{ms}$.  We can thus identify the breakdown of the
cluster expansion.  Note, however, that this is well into the tail of
the decay [where $v_E(\tau) < e^{-1000}$] and therefore this breakdown
is irrelevant for practical purposes.  It is prudent, in any case, to
understand the limitations of this expansion.

\begin{figure}
\includegraphics[width=3in]{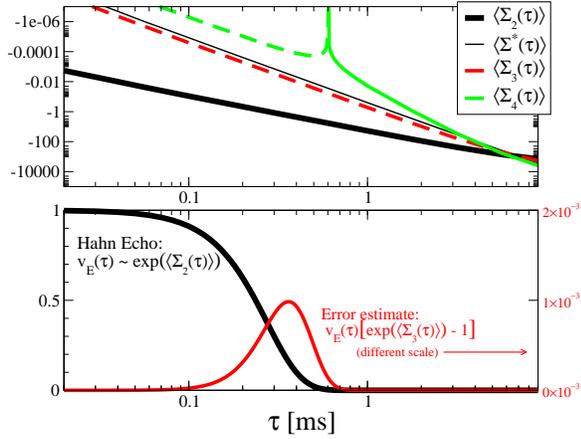}
\caption{
Low order corrections to the Hahn echo, $v_E(\tau)$, computed for Si:P in natural Si ($f
= 0.0467\%$) with $B~||~[100]$. 
(top) Log-log plot of low order contributions to the natural log of
the Hahn echo, $\ln{(v_E(\tau))}$.  Ordinate axis is negative as in
the bottom graph of Fig.~\ref{logSiHigherOrder}; however, dashed
lines indicate negated curves (and thus represent positive values).
(bottom) Conservative estimate of the absolute error of the lowest
order Hahn echo result
(scale on the right) due to $3$-cluster contributions,
$\left\langle\Sigma_3(\tau)\right\rangle$.  The lowest order Hahn
echo result is shown as a reference (scale on the left).  
The logarithmic time scale is the same for all plots (top and bottom).
\label{logNatSiHigherOrder}}
\end{figure}

The $\left\langle\Sigma_{k}(\tau)\right\rangle$ curves for some
fraction, $f$, of $^{29}$Si will be the same as the $\Sigma_{k}(\tau)$
curves in the log-log plots of Fig.~\ref{logSiHigherOrder} 
except with appropriate vertical shifts [multiplying by $f^k$
effectively appears as addition by $k \log{(f)}$ in the log plot] due
to the $f$ dependence.  Higher orders will be shifted closer to zero
than the lower order curves and therefore these curves will be more
separated (actually improving the cluster expansion convergence).  The
top graph of Fig.~\ref{logNatSiHigherOrder} is analogous to the bottom
(negative range) graph of Fig.~\ref{logSiHigherOrder} for natural Si
(dashed lines indicate negated curves, i.e., where values are actually positive).  
We show only the low order corrections to the log of the Hahn echo,
including $\left\langle\Sigma^{*}(\tau)\right\rangle$ as well as 
$\left\langle\Sigma_3(\tau)\right\rangle$ and
$\left\langle\Sigma_4(\tau)\right\rangle$ and not bothering with 
$\left\langle\Sigma_5(\tau)\right\rangle$.
$\left\langle\Sigma_3(\tau)\right\rangle$, with its inclusion of $3$-cluster, gives the largest
correction.  Although $\Sigma^{*}(\tau)$ is of a comparable order of
magnitude, its correction partially cancels the $\Sigma_3(\tau)$
correction because they are opposite in sign.  We may therefore use
$\Sigma_3(\tau)$ for a conservative estimate of the error of the
lowest order cluster expansion result.  The bottom graph of
Fig.~\ref{logNatSiHigherOrder} shows the absolute (as opposed to
relative) error of the lowest order Hahn echo result 
estimated by the inclusion of $\Sigma_3(\tau)$.  The
Hahn echo is displayed for reference.  At
its maximum, this absolute error is approximately $0.001$.  Although
our cluster expansion fails near $\tau \sim 5~\mbox{ms}$
[where $\lvert\left\langle\Sigma_{2}(\tau)\right\rangle\rvert \sim
\lvert\left\langle\Sigma^{*}(\tau)\right\rangle\rvert 
\sim \lvert\left\langle\Sigma_3(\tau)\right\rangle\rvert 
\sim \lvert\left\langle\Sigma_4(\tau)\right\rangle\rvert$], the absolute
error will stay small if we assume that our Hahn echo decay is forever 
monotonically
decreasing.  For all practical purposes, the
lowest order result is therefore valid up to $0.1\%$ of the initial
$v_E(0) = 1$, and higher order terms only provide corrections beyond
$99.9\%$ accuracy level.

\begin{figure}
\includegraphics[width=3in]{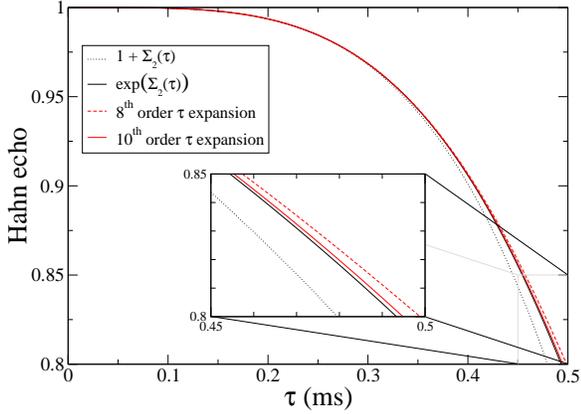}
\caption{
This compares the $\tau$-expansion results with the lowest order 
cluster-expansion 
result, $\exp{\left[\Sigma_2(\tau)\right]}$
[Eq.~(\ref{vE2_expFormApprox})],
 for an arbitrary set of nuclei 
(ranging from 15 to 20 lattice
constants away from the center) for the phosphorus donor in natural
silicon.  Shown are the $\tau$-expansion results at the highest two
orders of $\tau$ that we have computed.  Note the agreement with the
cluster expansion results before the two $\tau$-expansion results
diverge.  Also shown is $1 + \Sigma_2(\tau)$; its disagreement
indicates the importance of accounting for simultaneous pairs in the
cluster expansion.
\label{SiTauCluster}}
\end{figure}

We have also verified that our cluster expansion results agree
quantitatively with the $\tau$-expansion [Eq.~(\ref{vE_tau})] 
(applied directly, not in the context of the cluster expansion)
up to tenth order in $\tau$ 
when excluding
nuclei close to the center of the electron wave function where
$c_{nm}\gg 1$.  This provides an independent verification of the
cluster method approach.
An example is given in Fig.~\ref{SiTauCluster}.
As discussed in Sec.~\ref{dual_perturb_regimes}, however,
numerical calculations show that dipolar perturbation theory accounts
for the Hahn echo decay in physically relevant systems we have
studied.  
This is because clusters for which this
perturbation theory is applicable will tend to operate on shorter time
scales (due to the corresponding large hyperfine energies).  This has
been shown by comparing the lowest order cluster expansion 
using exact pair contributions versus approximate 
pair contributions in the lowest dipolar perturbation order
as discussed in Sec.~\ref{dual_perturb_regimes}.

\subsection{Gallium arsenide quantum dot}
\label{GaAsResults}

Next we consider the Hahn echo decay of a localized quantum dot
electron spin in GaAs.  For this, $\Psi(\bm{R}_n)$ will be the quantum dot wave
function parameterized by the quantum well thickness, $z_0$, and
Fock-Darwin radius, $\ell(B)$ (a function of the magnetic field
strength), as described in
Ref.~\onlinecite{desousa03b}. Using this in
Eq.~(\ref{An}), we have\cite{desousa03b}
\begin{eqnarray}
\label{An_GaAs}
A_n &=& \frac{16}{3} \frac{\gamma_{S} \gamma_{I} \hbar
  (a_{\mbox{\tiny GaAs}}^3 / 4)}{\ell^2(B) z_0} d(I) \cos^{2}{\left(\frac{\pi}{z_0} Z_n
  \right)} \\
\nonumber
&& \times \exp{\left(- \frac{X_{n}^{2} +
  Y_{n}^{2}}{\ell^{2}(B)}\right)} \Theta(z_0/2 - |Z_n|) \\
\nonumber
&& -\gamma_{S} \gamma_{I} \hbar \frac{1 - 3
  \cos^{2}{\theta_n}}{|\bm{R}_{n}|^{3}}
\Theta[X_n^2 + Y_n^2 - \ell^2(B)],
\end{eqnarray}
with $a_{\mbox{\tiny GaAs}} = 5.65$~\AA~and 
$\gamma_{S} = 1.76 \times 10^7 \mbox{(s G)$^{-1}$}$ (the free electron
gyromagnetic ratio).
The GaAs lattice has a zinc-blende structure with two isotopes of Ga atoms
placed on one fcc lattice and $^{75}$As atoms placed on the other fcc
lattice.\cite{kittel}  
The Ga isotopes are $60.4\%$ $^{69}$Ga and $30.2\%$ $^{71}$Ga.\cite{desousa03b,
  chem_handbook}  We used
$\gamma_{I} = 4.58, 8.16, 6.42 \times 10^3 \mbox{(s G)$^{-1}$}$ 
and $d(I) = 9.8, 5.8, 5.8 \times 10^{25}~\mbox{cm}^{-3}$
for
$^{75}$As, $^{71}$Ga, and $^{69}$Ga, respectively.\cite{desousa03b,
  paget77}
All of these nuclei have a valence spin of
$I = 3/2$ which means that Eq.~(\ref{vnm}) is not applicable.
Instead, cluster contributions are computed by numerically diagonalizing
${\cal H}^{\pm}$.

Most of our results only
include dipolar interactions for $b_{nm}$ [Eq.~(\ref{bnm})].
However, indirect exchange interactions
\cite{shulman58, bloembergen55, sundfors69} 
between the nuclei of GaAs may be of the same order of
magnitude as the dipolar interactions [Eq.~(\ref{bnm})]
between nearest neighbors.\cite{yao05}  There is enough
quantitative ambiguity in the literature about the indirect exchange
interaction that we have chosen to leave it out of our calculations in
order to have a precise theory for the dipolar nuclear spin bath
dynamics only.
We make an exception near the end of this section where we make
a comparison to the results in Ref.~\onlinecite{yao05} that include this
interaction. 

With exception to the $I_{nz} I_{mz}$ term in ${\cal H}^B$ 
[Eq.~(\ref{H_Bnm})],
the different types of nuclei are decoupled.  In the lowest order of
the cluster expansion, the different types of nuclei truly are
decoupled because a pair cannot contribute to the spectral diffusion
without the flip-flop interaction between them.  Furthermore, this
$b_{nm} I_{nz} I_{mz}$ term is negligible where the dipolar
perturbation is applicable because the diagonal of the Hamiltonian in the nuclear
$z$-basis will be dominated by ${\cal H}^A$.  
We will later discuss comparisons made between calculations using 
exact pair contributions versus approximate pair contributions
using the lowest order dipolar perturbation (also done in
Sec.~\ref{Si:P} for Si:P) to demonstrate that the dipolar perturbation
theory is applicable.  Using these arguments, we will simplify our
calculations by approximating Eq.~(\ref{H_Bnm}) with
\begin{equation}
\label{HB_flipflop_only}
{\cal H}^{\cal B} \approx \sum_{n \ne m} b_{nm} \delta(\gamma_n -
\gamma_m) I_{n+} I_{m-}.
\end{equation}
This approximation effectively decouples
the As nuclei from the Ga nuclei and the Ga isotopes from each other.
Although it is possible to perform our cluster calculations without
this approximation (we have done so for some lowest order calculations
with no additional complication and the difference is indeed
negligible), it avoids unnecessary 
complications for higher order calculations.
We may then perform calculations for each of these nuclear isotopes
separately.
Using the subscript $x$ to represent each of these types of nuclei,
the contributions to the natural log of the Hahn echo become
\begin{eqnarray}
\left\langle\Sigma_k(\tau)\right\rangle = \sum_x f_x^2
\Sigma_{kx}(\tau), \\
\left\langle\Sigma_k^{*}(\tau)\right\rangle = \sum_x f_x^2
\Sigma_{kx}^{*}(\tau),
\end{eqnarray}
where $\Sigma_{kx}(\tau)$,  $\Sigma_{2x}^{*}(\tau)$, and
$\Sigma_{3x}^{*}(\tau)$ [Eqs.~(\ref{Sigma_j}), (\ref{sigma2_star}), and
(\ref{sigma3_star}), respectively] sum over clusters in the
appropriate sublattice of isotope $x$, 
and $f_x$ is the fraction of this isotope in
its sublattice.
For $^{75}$As, $^{71}$Ga, and $^{69}$Ga, it
is appropriate to use $f_x = 100\%$,~$30.2\%$,~and~$60.4\%$,
respectively.  Monte Carlo techniques (described in Sec.~\ref{Si:P}) 
are used to approximate these sums.

The lowest order results, $v_E(\tau) \approx
\exp{\left[\left\langle\Sigma_2(\tau)\right\rangle\right]}$, 
for most of our GaAs calculations show a Hahn echo 
decay of the form $\exp{\left[-(2 \tau/t_0)^4\right]}$.  This differs
qualitatively from the decay for the Si:P which, by
our calculations, is in the form
$\exp{\left[-(2 \tau/t_0)^\alpha\right]}$ where $\alpha \sim 2.3$ for
a range of $\tau$ appropriate for natural Si and some range of
isotopic purification.
The form of the GaAs echo decay does not change if we repeat the
calculation with $I=1/2$ rather than $3/2$, although $t_0$ does
change.  The qualitative difference between Si:P and GaAs dot
solutions is therefore not due to the difference in spin magnitude of
the nuclei, but rather the difference in the form of the electron wave 
function and lattice occupation leading to different distributions of $c_{nm}$ values.
GaAs, in ranges of tested $z_0$ and $\ell$ parameters, has a more
narrow distribution of $c_{nm}$ for significantly contributing pairs
than that of the Si:P problem.  As a result of this narrow distribution, 
the fact that the lowest nontrivial order of
the $\tau$-expansion is ${\cal O}(\tau^4)$, which is universal for all values
of $I$, expresses itself in the sum of pair contributions in
 $\Sigma_2(\tau)$ [Eq.~(\ref{Sigma_j})].  This 
$\exp{\left[-(2 \tau/t_0)^4\right]}$ form is confirmed by Yao 
{\it et al.}\cite{yao05} using a completely different theoretical technique.  
Yao {\it et al.} use a pair-correlation
quasiparticle approach to the ``spectral diffusion'' problem that
is equivalent to our lowest order cluster expansion, and they
give results for quantum dots in GaAs.  They come to the same conclusion
that the form of the decay depends on the relative breadth of the
excitation spectrum.

\begin{figure}
\includegraphics[width=3in]{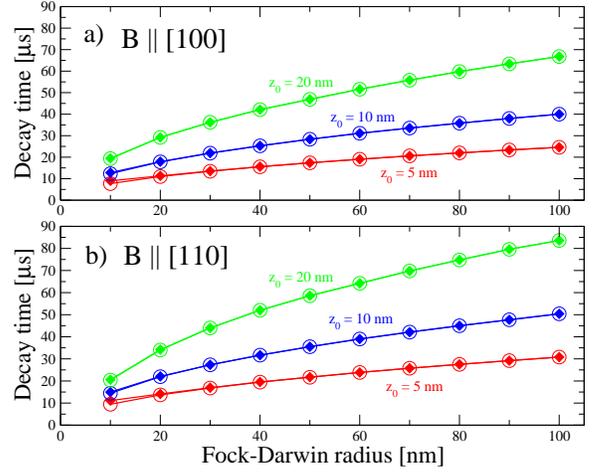}
\caption{
For GaAs quantum dots,
$t_0$ (circles), the characteristic initial decay time, 
and $t_{1/e}$ (diamonds), 
the $e^{-1}$ decay time,
vs the Fock-Darwin radius $\ell$ for various
quantum well thicknesses, $z_0 = 5, 10,~\mbox{and}~20$~nm.  
The orientation of the magnetic field is (a)
along the $z_0$ confinement of the quantum dot and [100] lattice
direction, or (b) perpindicular to the $z_0$ confinement direction and
along [110] of the lattice.
\label{GaAsB0B90combined}}
\end{figure}

Figure~\ref{GaAsB0B90combined}
 shows the $t_0$ of the initial 
$\exp{\left[-(2 \tau/t_0)^4\right]}$ Hahn echo decay for various
parameter settings of $z_0$ and $\ell$ with two different magnetic
 field orientations.  Also shown is $t_{1/e}$, defined such that
$v_E(\tau = t_{1/e}/2) = e^{-1}$.  One can think of this $t_{1/e}$ as
 an effective $T_2$-time for the problem although the echo decay is
 not a simple exponential.  Except for small dots, $t_0 =
 t_{1/e}$, indicating that the decay has the form
$\exp{\left[-(2 \tau/\tau_0)^4\right]}$.  Small dots deviate from this
 form, beginning to have longer $t_{1/e}$ decay times than their
 initial characteristic times, $t_0$.  It was noted in
Ref.~\onlinecite{desousa03b} that decoherence times become infinite 
as the size of the quantum dot approaches zero or infinity with
a minimum decoherence time at some finite size.  The former is simply because the
 electron has no interaction with nuclei as the quantum dot size
 approaches zero, and the latter is because the nuclei all have the
 same coupling to the electron as the size becomes infinite.
For $z_0 = 5~$nm we begin to approach this maximum decoherence
 (minimum $t_{1/e}$) near $\ell = 10~$nm, but only in the regime where 
$t_{1/e}$ deviates from $t_0$.

As discussed previously, the Ga and the As nuclei, on separate fcc
lattices, are decoupled by our approximation [Eq.~(\ref{HB_flipflop_only}).
In silicon, the asymmetry of
the diamond lattice results in maximum decoherence in the $[111]$
direction.  In this case, because the fcc lattice is more symmetric,
the angular dependence is primarily a result of the shape of the
quantum dot (not the lattice).  Figure~\ref{GaAsB0B90combined}
shows slight quantitative differences when the magnetic field is along
the $z_0$ confinement direction or perpendicular to it.

Because most of our GaAs results are in the form corresponding to the
limit of small $\tau$, it is tempting to think that GaAs is dominated
by the $c_{nm} \ll 1$ regime appropriate for the $\tau$-expansion.
However, as with the phosphorus donor in Si, we have compared 
calculations that use exact pair contributions versus approximate 
pair contributions using the lowest order dipolar perturbation (also
discussed in Sec.~\ref{dual_perturb_regimes}).
These different calculations agree very well for small quantum dots,
but deviate slightly for larger quantum dots.  Intermediate sized dots
are well-approximated by either perturbation theory.  This is merely a result
of small decoherence times such that the $\tau$ approximation is valid
even though the dominating pairs have $c_{nm} > 1$.  The agreement with the
perturbation expansion gives justification for the approximation used in
Eq.~(\ref{HB_flipflop_only}) in which $b_{nm} I_{nz} I_{mz}$ was
neglected.

\begin{figure}
\includegraphics[width=3in]{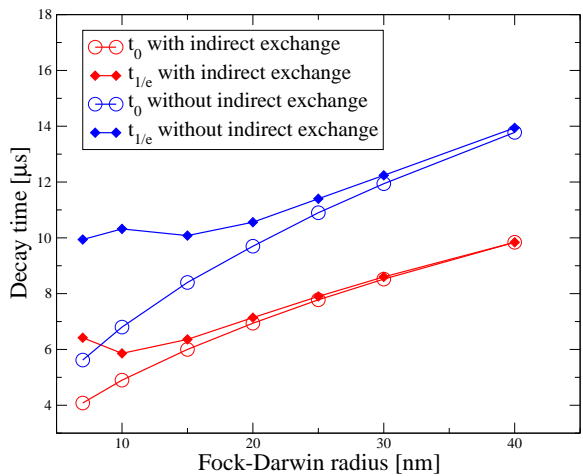}
\caption{
For GaAs quantum dots, $t_0$, the characteristic initial decay
time, and $t_{1/e}$, the $e^{-1}$ decay time,
vs the Fock-Darwin radius $\ell$ for a
quantum well thicknesses of $z_0 = 2.8$~nm.  
The orientation of the magnetic field is parallel to $[110]$ of the
lattice; this is perpendicular to the $z_0$ confinement of the quantum
dot.  This shows results both when including or excluding the indirect
exchange coupling.
\label{yao_compare}}
\end{figure}

We now return to a discussion of the indirect exchange interaction
between nuclear spins (mediated by virtual interband electronic
transitions) that were neglected in the above calculations.  
Including the exchange interaction, we should use
\begin{equation}
b_{nm} = b_{nm}^{d} + b_{nm}^{e},
\end{equation}
where $b_{nm}^{d}$ is the dipolar coupling [Eq.~(\ref{bnm})], and 
$b_{nm}^{e}$ is the indirect exchange coupling.  
We note that $b_{nm}^{e} = 0$ in the Si:P system to a high degree of accuracy.
Yao {\it et al.}\cite{yao05} performed spectral
diffusion decoherence calculations (using an equation that is equivalent to
our lowest order result) for GaAs quantum dots including the indirect
exchange interaction.  As a verification of the correctness
of our calculations, Fig.~\ref{yao_compare} reproduces
their Hahn echo results using our method but including the 
indirect exchange interactions as given by\cite{sundfors69}
\begin{equation}
\label{bnm_e}
b_{nm}^{e} = 
-\gamma_{n}\gamma_{m} \hbar\frac{\sqrt{2.6}~\mbox{\AA}}{2 R_{nm}^4}.
\end{equation}
There is ambiguity in the literature about the appropriate sign of
  Eq.~(\ref{bnm_e}).  We have adopted
the convention used by Yao {\it et al.}\cite{yao05} in order to
reproduce their results.  Figure~\ref{yao_compare} also shows the
results for the same parameters when the indirect exchange is
excluded; it is apparent that this coupling is significant, at least
  in GaAs quantum dots for these choices of the exchange coupling.  The kink
in the $t_{1/e}$ curve for the case of excluded indirect exchange is
believed to be a discrete lattice effect only noticeable for small
quantum dots.  For such small quantum dots, it is likely that
Eq.~(\ref{An_GaAs}), derived from an approximate electron wave
function, is somewhat inaccurate.

\begin{figure}
\includegraphics[width=3in]{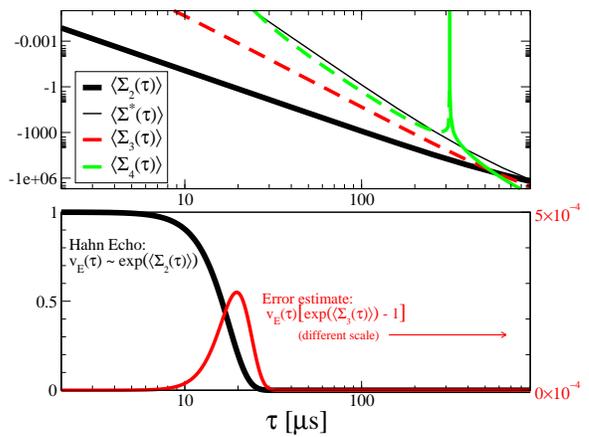}
\caption{
Low order corrections to the Hahn echo, $v_E(\tau)$, 
computed for a GaAs quantum
dot with $B~||~[110]$,  $z_0 = 10~\mbox{nm}$, and $\ell =
50~\mbox{nm}$. 
(top) Log-log plot of low order contributions to the natural log of
the Hahn echo, $\ln{(v_E(\tau))}$.  Ordinate axis is negative; 
however, dashed lines indicate negated curves 
(and thus represent positive values).
(bottom) Conservative estimate of the absolute error of the lowest
order Hahn echo result
(scale on the right) due to $3$-cluster contributions,
$\left\langle\Sigma_3(\tau)\right\rangle$.  The lowest order Hahn
echo result is shown as a reference (scale on the left).  
The logarithmic time scale is the same for all plots (top and bottom).
\label{logGaAsHigherOrder}}
\end{figure}

With our cluster expansion approach, we can estimate the error of
our calculated decay curves by performing higher order calculations.  We
observe that the larger quantum dots have larger corrections.
For quantum dots with $z_0 = 20$~nm and $\ell = 100$~nm our
calculations indicate maximum correction to the Hahn echo decay curves
on the order of $10^{-3}$, $0.1\%$ of the initial $v_E(0)=1$, just as it was for natural silicon 
(Fig.~\ref{logNatSiHigherOrder}).  For dots
with $z_0 = 5$~nm and $\ell = 10$~nm, absolute corrections are on the order
of $10^{-4}$, $0.01\%$ of the initial $v_E(0)=1$.  
Fig.~\ref{logGaAsHigherOrder}, analogous to
Fig.~\ref{logNatSiHigherOrder}, 
gives these low order corrections explicitly
for an intermediate size 
($z_0 = 10~\mbox{nm}, \ell = 50~\mbox{nm}$) GaAs quantum dot.

\section{Conclusion}
\label{conclusion}
In conclusion, we describe a quantum approach for the problem of localized
electron spin decoherence due to the fluctuation of dipolar
coupled nuclear spins. In contrast to former theories, our method
requires no {\it ad hoc} stochastic assumption on the complex dynamics
of the environment responsible for decoherence. Hence it provides an
important example where direct integration of the environmental
equations of motion provides a systematic understanding of the loss of
coherence which needs to be controlled for quantum information
applications. 

The most important theoretical accomplishment of our work is the
development of the first {\it fully quantum microscopic} theory
for the localized electron spin decoherence due to the spectral
diffusion induced by nuclear spin bath dynamics.  The important
nuclear spin dynamics for the spectral diffusion problem is the
dipolar interaction induced nuclear spin flip-flops although for the
GaAs quantum dot system, we have also included the indirect exchange
interaction between the nuclear spins which turns out to be
quantitatively comparable to the dipolar contribution to spectral
diffusion.  Our results are formally exact, and our numerical
calculations provide an
essentially exact (to better than $0.1\%$ of the initial value) quantitative description of
the Hahn spin echo decay.  The significance of
our quantum theory lies in the fact that, unlike all other theoretical
descriptions of spectral diffusion spanning the last $50$ years, we
do not make any {\it ad hoc} phenomenological stochastic
approximation in dealing with the non-Markovian spin dynamics in the
spectral diffusion phenomena.  We solve the problem essentially
exactly using a quantum cluster decomposition technique, which is then
theoretically justified by carrying out calculations to higher orders
and by comparing with two independent perturbation techniques
(i.e., the tau expansion and the dipolar perturbation theory).
Very recently, a completely independent verification of our theory and
results has appeared in the literature.\cite{yao05}

Our theory for spin decoherence is restricted to
understanding the so-called $T_2$-decoherence of a single localized
spin in a solid 
due to the dipolar nuclear spin flip-flops in the
surrounding nuclear spin bath
and in the specific context of a spin echo experiment.
We neglect spin decoherence effects arising from the direct
hyperfine coupling between the central spin and the surrounding
nuclear spins, which would contribute to the free induction decay
signal, but can be corrected by the spin echo technique.
Our method uses a cluster expansion technique supported by two
underlying perturbation theories:
the dipolar perturbation expansion and the tau-expansion.
Each of these perturbation theories generates
a unique expansion scheme in its own right, but only the cluster
expansion converges for large systems
[only for short times, $\tau \ll \Max{(b_{nm})}^{-1}$, but often extending to the tail of the decoherence decay].
The convergence of this expansion has to be explicitly checked in each
application, as we have done in this work.  It is possible that an
alternative more conventional perturbation theory for the
non-Markovian spectral diffusion problem can be developed within the
standard interaction picture of the many-body quantum theory which
would be completely equivalent (for short times) to our cluster
expansion technique.  More work would be needed to clarify and
establish such a possibility.

We note that spectral diffusion induced decoherence cannot be
characterized by a simple $T_2$ time since the Hahn echo decay is not
a simple exponential, and in fact, obeys different temporal power laws
in the Si:P and the GaAs quantum dot systems.  Spectral diffusion
is a pure dephasing process arising from the temporal
(non-Markovian) magnetic field fluctuations at the localized electron
spin due to the nuclear spin dynamics, but it cannot be parameterized
by a single $T_2$ time except as a crude approximation.  Within this
crude approximation, we find the spectral diffusion induced $T_2$ to
be about $100~\mbox{$\mu$s}$ for the Si:P system and about
$10~\mbox{$\mu$s}$ for the GaAs quantum dot system.  But, this $T_2$
can be enhanced indefinitely (up to tens of milliseconds) in the Si:P
system through the isotopic purification of Si (i.e., by removing
$^{29}$Si nuclei from the system) whereas in the GaAs quantum dots,
$T_2 \sim 10~\mbox{$\mu$s}$ is essentially an absolute upper limit (when
using simple Hahn echo refocusing) since {\it all} 
Ga and As nuclei isotopes have free
spins contributing to the spectral diffusion and isotopic purification
is impossible.  It is important to emphasize here that although spin
polarizing the nuclei (e.g., through the dynamic nuclear polarization
technique) would, in principle, suppress the nuclear induced spectral
diffusion decoherence of electron spin, in practice, this would lead
only to rather small enhancement of electron spin coherence since the
presence of even a few nuclei with the ``wrong'' spin would cause
nuclear pair flip-flop processes.\cite{SDS05}

Finally, we comment on the fact that the spectral diffusion process is
quite a generic and general phenomenon in {\it any} spin
decoherence problem with coupled spin dynamics (e.g. electron and
nuclear spins, different types of nuclear spins, etc.) where the
dynamics of one spin species has nontrivial (i.e., non-Markovian)
temporal effects on the evolution of the spin dynamics of the other
species.  For example, a trivial (but not often emphasized in the
literature) consequence of spectral diffusion consideration implies
that in systems (e.g., Si:P; GaAs quantum dots) of interest to quantum
computer architectures, the single flip of the localized electron spin
will immediately decohere {\it all} the nuclear spins in its
vicinity.  Thus the nuclear spin $T_2$ time in these systems can at
most be the $T_1$ time for the electron spin!  The typical
low-temperature $T_1$ time for electron spins in the GaAs quantum dots
has been measured to be $1~\mbox{ms}$ or so, and therefore the nucleus
spin $T_2$ time would at most be $1~\mbox{ms}$ in the GaAs quantum
dots, at least in the neighborhood of the localized electrons in the
dot.  The same consideration applies to the Si:P system.  We believe
that the general quantum theoretical techniques developed in this
paper will be helpful in the studies of the temporal dynamics of other
coupled spin systems wherever one spin species could act as a
``decoherence bath'' for the other system.

\section{Acknowledgments}
We wish to thank Rogerio de Sousa, Eisuke Abe, Alexei
Tyryshkin, Ren-Bao Liu, and most particularly Wang Yao for valuable 
correspondence and discussion.
This work is supported by ARO-ARDA and LPS-NSA.

\appendix
\section{Hahn echo analysis}
\label{appendix_analysis}

Here we derive Eq.~(\ref{v_E_temp}) from Eq.~(\ref{amplitude_def}) and
related equations.  By noting that
\begin{equation}
S_{x} + i S_{y} = S_{+} = |1_e\rangle\langle0_e|,
\end{equation}
we may rewrite Eq.~(\ref{amplitude_def}),
\begin{equation}
\label{amplitude_def_appendix}
v_{E}(\tau) = 2~\Tr_{n}({\langle0_e|\rho(\tau)|1_e\rangle}), 
\end{equation}
where the $n$ subscript in $\Tr_{n}{}$ denotes a trace over the
nuclear subspace.  In addition to this equation, we will be referring to 
Eqs.~(\ref{H})-(\ref{H_Zn}), 
(\ref{H_A}), and (\ref{H_Bnm}) for the free evolution Hamiltonian,
${\cal H} = {\cal H}^{Ze} + {\cal H}^{Zn} + {\cal H}^{A} + {\cal
  H}^{B}$, as
well as Eqs.~(\ref{rho_t})-(\ref{chi_0}) to define the density
matrix, $\rho(\tau)$, in terms of the evolution operator, $U(\tau)$,
and inital density matrix, $\rho_{0}(\tau)$.

First we note that ${\cal H}^{Ze}$ commutes with the rest of the
Hamiltonian and that ${\cal H}^{Zn}$ commutes with ${\cal H}^{A}$
(since an operator commutes with itself and
any operator that acts on an orthogonal state space).
${\cal H}^{Zn}$ also commutes with ${\cal H}^{B}$ (and therefore commutes with the
entire Hamiltonian).  This is proven by noting that ${\cal H}^{B}$ preserves the
overall polarization of like nuclear spins.  That is,
\begin{equation}
\label{polarization_conserved}
\left[\gamma_{n} I_{nz} + \gamma_{m} I_{mz}, \delta{(\gamma_n - \gamma_m)}
  I_{n+} I_{m-}\right] = 0.
\end{equation}

Let us define
\begin{equation}
\label{U_0}
U_0(t) = e^{-\imaginary {\cal H} t},
\end{equation}
and exploit the above commutation relations to write
\begin{eqnarray}
U_{0}(t) & = & e^{-\imaginary {\cal H}^{Zn} t} e^{-\imaginary {\cal H}^{Ze} t}
U'_{0}(t), \\
\label{U'_0_appendix}
U'_{0}(t) & = & e^{-\imaginary ({\cal H}^{A} + {\cal H}^{B}) t}.
\end{eqnarray}

Using
\begin{equation}
\left\{\sigma_{x, e}, S_z\right\} = 0 \Rightarrow \sigma_{x, e} e^{-\imaginary {\cal H}^{Ze} \tau} = e^{i {\cal H}^{Ze} \tau} \sigma_{x, e},
\end{equation}
and the fact that ${\cal H}^{Zn}$ commutes with $\sigma_{x, e}$ (the former
acting only on the nuclear spins and the latter acting on the electron
spin), we can rewrite Eq.~(\ref{U})
\begin{eqnarray}
U(\tau) & = & U_0(\tau) \sigma_{x, e} U_0(\tau) \\
        &=& U'_{0}(\tau) \sigma_{x, e} U'_{0}(\tau) e^{-\imaginary {\cal H}^{Zn} (2 \tau)}\\
        & = & U'(\tau) e^{-\imaginary {\cal H}^{Zn} (2 \tau)},\\
\label{U'_appendix}
U'(\tau) & = & U'_{0}(\tau) \sigma_{x, e} U'_{0}(\tau).
\end{eqnarray}

We can then rewrite Eq.~(\ref{rho_t})
\begin{eqnarray}
\rho(\tau) & = & U'(\tau) e^{-\imaginary {\cal H}^{Zn} (2 \tau)}\rho_{0} e^{i {\cal H}^{Zn} (2 \tau)}
U'^{\dag}(\tau) \\
\label{rho'}
              & = & U'(\tau) \rho_{0} U'^{\dag}(\tau),
\end{eqnarray}
using the fact that ${\cal H}^{Zn}$ commutes with $\rho_{0}$ (since it
commutes with 
$\left|\chi_{e}^{0}\right>\left<\chi_{e}^{0}\right|$, and ${\cal H}^n =
{\cal H}^{Z_n} + {\cal H}^B$).

At this point it is convenient, for clarity, to write our operators as block
matrices in the electron spin $z$ basis.  First we have, from Eq.~(\ref{U'_0_appendix})
\begin{eqnarray}
\label{U'_0_block_appendix}
U'_{0}(\tau) & = & \left[\begin{array}{cc}
U_{+} & 0 \\
0 & U_{-}
\end{array}\right],
\end{eqnarray}
where
\begin{equation}
\label{Upm_appendix}
U_{\pm} = e^{-\imaginary {\cal H}^{\pm} \tau};~U_{\pm}^{\dag} = e^{i {\cal
    H}_{\pm} \tau},
\end{equation}
with
\begin{eqnarray}
\label{Hpm_appendix}
{\cal H}^{\pm} & = & {\cal H}^{B} \pm {\cal H}^{An}, \\
{\cal H}^{An} & = & \frac{1}{2} \sum_{n} A_{n} I_{nz}.
\end{eqnarray}
$U_{\pm}$ can be interpreted as the evolution operators of the nuclei with the
electron spin constrained to be up or down, respectively, 
and ignoring the external magnetic field.  
Although $U_{+}$ and $U_{-}$ are functions of $\tau$,
we have dropped this as an explicit parameter for convenience.

In this same electron spin basis, we can write
\begin{equation}
\begin{array}{cc}
  \left|\chi_{e}^{0}\right>\left<\chi_{e}^{0}\right| = \frac{1}{2} \left[\begin{array}{cc}
      1 & e^{i \phi} \\
      e^{-\imaginary \phi} & 1
    \end{array}\right], &
  \sigma_{x, e} = \left[\begin{array}{cc}
      0 & 1 \\
      1 & 0
    \end{array}\right],
\end{array}
\end{equation}
in order to obtain, from Eqs.~(\ref{rho_0}), (\ref{rho'}), (\ref{U'_appendix}), and (\ref{U'_0_block_appendix}),
\begin{equation}
\langle0_e|\rho(\tau)|1_e\rangle = \frac{e^{\imaginary \phi}}{2 M} 
 U_{-} U_{+} e^{-{\cal H}^n/T}
U_{-}^{\dag} U_{+}^{\dag}.
\end{equation}

Plugging this into Eq.~(\ref{amplitude_def_appendix}) gives
\begin{equation}
\label{vE_appendix}
v_{E}(\tau)=\frac{e^{\imaginary \phi}}{M} \Tr{\left\{
U_{-}U_{+}e^{-{\cal H}^n/k_B T}
U_{-}^{\dag}U_{+}^{\dag}
\right\}}.
\end{equation}
We are interested in the echo envelope which is given by the 
magnitude of Eq.~(\ref{vE_appendix}) and
thus the $e^{\imaginary \phi}$ phase factor may be dropped.

\section{Independence of Cluster Contributions}
\label{appendix_clusterIndependence}

In order to prove the unique existence of the decomposition of 
$v_{\cal S}(\tau)$ [the solution to the Hahn echo, $v_E(\tau)$, when
only including the nuclei in some set ${\cal S}$] given by 
Eq.~(\ref{vS_decomposition}), we must prove that each
cluster contribution is independent of anything outside of that
cluster.
Only then can we assume that the contribution of a given cluster is
the same in every instance (i.e., in every term that involves this
cluster) in such a decomposition.

Let us consider some cluster ${\cal C}$ contained in ${\cal S}$.  
Because we are only concerned with the independence of processes in 
disjoint sets, let us remove any processes from $v_{\cal S}(\tau)$ 
with coupling between nuclei 
in ${\cal C}$ and nuclei in ${\cal S} - {\cal C}$.  
Define $v_{{\cal A} : {\cal B}}(\tau)$ to be the sum of all of the terms 
in the sum of products expansion (described at the beginning of
Sec.~\ref{clusterdefinition}) of 
$v_{({\cal A} \bigcup {\cal B})}(\tau)$ containing any factor of 
$b_{nm}$ in which $n \in {\cal A}$ and $m \in {\cal B}$ (or
vice versa).  Thus, by definition,
$v_{\cal S}(\tau) - v_{{\cal C} : ({\cal S} - {\cal C})}(\tau)$ does 
{\it not} involve {\it any} coupling between a nucleus in ${\cal C}$
and a nucleus in ${\cal S} - {\cal C}$; it only involves coupling
amongst nuclei in ${\cal C}$ or amongst nuclei in ${\cal S} - {\cal
  C}$.  This is the same as calculating $v_{\cal S}(\tau)$ if we
artificially impose the condition that
\begin{equation}
\label{imposeDisjoint}
b_{nm} = b_{mn} = 0 ~\forall~ n \in {\cal C},~  m \in {\cal S} - {\cal C}.
\end{equation}

For convenience, we define the following:
\begin{eqnarray}
V_{\cal S} &=& U_{{\cal S}-} U_{{\cal S}+} U_{{\cal S}-}^{\dag} 
U_{{\cal S}+}^{\dag}, \\
U_{{\cal S}\pm} &=& e^{-\imaginary {\cal H}^{\pm}_{\cal S}\tau},\\
{\cal H}^{\pm}_{\cal S} &=& 
\sum_{
\substack{
n \ne m \\
n, m \in {\cal S}
}} {\cal H}^{B}_{nm}
\pm \frac{1}{2} \sum_{n \in {\cal S}} A_n I_{nz},
\end{eqnarray}
such that
\begin{equation}
v_{\cal S}(\tau) = \frac{1}{M}\Tr{\left\{V_{\cal
    S}\right\}}
= \frac{1}{M_{\cal S}} \Tr_{\cal S}{\left\{V_{\cal S}\right\}},
\end{equation}
where $M_{\cal S}$ is the number of composite states for the nuclei in
set ${\cal S}$ and $\Tr_{\cal S}$ takes the trace over only these states.
Note that if we artificially impose condition (\ref{imposeDisjoint}), then
\begin{equation}
{\cal H}^{\pm}_{\cal S} \rightarrow {\cal H}^{\pm}_{\cal C} + 
{\cal H}^{\pm}_{({\cal S} - {\cal C})}.
\end{equation}
Noting that ${\cal H}^{\pm}_{\cal C}$ commutes with
${\cal H}^{\pm}_{({\cal S} - {\cal C})}$ (since they act on disjoint
sets of nuclei), then
\begin{eqnarray}
U_{{\cal S}\pm} &\rightarrow& e^{-\imaginary {\cal H}^{\pm}_{\cal C}\tau}
e^{-\imaginary {\cal H}^{\pm}_{({\cal S} - {\cal C})} \tau}\\
 &=& U_{{\cal C}\pm} U_{({\cal S} - {\cal C})\pm}, \\
V_{\cal S} &\rightarrow& V_{\cal C} V_{({\cal S} - {\cal C})}.
\end{eqnarray}
We may therefore write
\begin{eqnarray}
v_{\cal S}(\tau) - v_{{\cal C} : ({\cal S} - {\cal C})}(\tau) &=& 
\frac{1}{M_{\cal S}}\Tr_{\cal S}{\left\{V_{\cal C} V_{({\cal S} - {\cal
      C})}\right\}} \\
\label{TraceFactoringV}
&=& \frac{ \Tr_{\cal C}{\left\{V_{\cal C}\right\}}
\Tr_{({\cal S} - {\cal C})}{\left\{V_{({\cal S} - {\cal C})}\right\}}}
{M_{\cal C} M_{({\cal S} - {\cal C})}}~~~~ \\
&=& v_{\cal C}(\tau) v_{({\cal S} - {\cal C})}(\tau).
\end{eqnarray}
Equation~(\ref{TraceFactoringV}) is a direct consequence of the fact
that $V_{\cal C}$ and $V_{({\cal S} - {\cal C})}$ act on orthogonal
subspaces of the Hilbert space because they act on disjoint sets of nuclei.
We conclude that
\begin{equation}
\label{disjointIndependence}
v_{\cal S}(\tau) = v_{\cal C}(\tau) v_{({\cal S} - {\cal C})}(\tau)
+ v_{{\cal C} : ({\cal S} - {\cal C})}(\tau),
\end{equation}
proving that processes involving disjoint sets of nuclei are
independent for the following reason.  Any interdependent process of
cluster ${\cal C}$ must be contained in the $v_{\cal C}(\tau)$ part of
the right-hand side of Eq.~(\ref{disjointIndependence}).  This has no
dependence on the set ${\cal S}$ beyond the requirement that 
${\cal C}$ be contained in it.  For this reason, the cluster
contribution for cluster ${\cal C}$ is well-defined in the sense that
it is independent of all other nuclei.

We may also verify from
Eq.~(\ref{disjointIndependence}) that we are to multiply independent
cluster contributions in order to account for simultaneous processes 
[the first term on the right side of Eq.~(\ref{disjointIndependence})]
and that alternative combinations are to be included via addition
(the second term).  With this observation, we can
decompose any $v_{\cal S}(\tau)$ into a sum of products of
cluster contributions.  
Recall that a cluster contribution
includes all contributions from interdependent processes involving
{\it only} and {\it all} of the nuclei in the cluster.  
Each process is thus assigned to only one cluster and  
there is no redundancy or overlap between different cluster contributions.
Therefore, without redundancy, $v_{\cal S}(\tau)$ is the sum of {\it all
  possible} products of cluster contributions as given by 
Eq.~(\ref{vS_decomposition}).

\section{Methods for Computing the Algebraic Tau-Expansion}
\label{appendix_ComputerAlgebra}

We describe techniques used in our computer program for
algebraically expanding  Eq.~(\ref{v_E}) in orders of $\tau$.
For a particular order of $\tau$, an expansion of the exponentials
of $U_{\pm}(\tau)$ [Eq.~(\ref{Upm})]
will result in the trace of a sum of products of
${\cal H}^{\pm}$, and ${\cal H}^B$.  The ${\cal
  H}_{\pm}$ factors [Eq.~(\ref{Hpm})] can all be
distributed to yield a sum of traces of products of ${\cal H}^B$,
and ${\cal H}^{An}=\frac{1}{2} \sum_{n} A_{n}
I_{nz}$.  Each one of the Hamiltonian pieces, ${\cal H}^{An}$, and ${\cal H}^B$,
involves sums over nuclear site indices.  We can decrease the
number of independent indices by one (not a major improvement,
but enough to push this method one step further) by noting that

\begin{equation}
\label{AB_commutation}
\left[{\cal H}^B, {\cal H}^{An}\right] = \frac{1}{2} \sum_{n \ne m} b_{nm} (A_n -
A_m) I_{n-} I_{m+},
\end{equation}
which takes away the independence of ${\cal H}^{An}$'s index.  Let ${\cal H}^C = \left[{\cal H}^B,
  {\cal H}^{An}\right]$.  We can convert our sum of traced products of
  ${\cal H}^{An}$ and
${\cal H}^B$ into a sum of traced products of ${\cal H}^{An}$ and
${\cal H}^B$ and a single ${\cal H}^C$ with one less independent summation index.
The rest of this paragraph will briefly explain how this is done.
Consider all of the terms that involve a certain number of 
${\cal H}^{B}$'s and ${\cal H}^{An}$'s.  
For
  any term, we can permute the ${\cal H}^{An}$'s and ${\cal H}^{B}$'s however we
want at the expense of creating terms in which an ${\cal H}^{B} {\cal
  H}_{An}$ combination is replaced by ${\cal H}^C$ 
(which is the type of term we want anyway).
Consider the set of terms that all have a certain number of ${\cal
  H}_{An}$ factors and a certain number of ${\cal H}^{B}$ factors.
As long as the coefficients of such a set add up to zero,
we can make the permutations cancel out, leaving only the terms with
${\cal H}^C$ and one less independent summation index.  By the symmetry in $e^{\pm i {\cal H}^{\pm}
  \tau}$ of Eqs.~(\ref{v_E}) and (\ref{Upm}),
these coefficients add up to zero for all such sets in any order of $\tau$.

Consider pulling the summations over nuclear site indices, not only outside of each product, 
but also outside of the trace 
(since the trace of a sum is equal to the sum of the traces).
The trace of these spin operators involving particular nuclear sites
is straightforward to compute.  In fact, all one really needs to know
to perform the trace is which indices are the same and which ones are
different (as well as the magnitude of the nuclear spins).  Then one can compute the trace as follows.  
If $O_n$ is some operator (a product of spin operators $I_{nz}$,
$I_{n+}$, $I_{n-}$) that operates on the
spin of site $n$, then
\begin{equation}
  \frac{1}{M} \Tr{(O_1 O_2 ... O_k)} = \frac{\Tr{(O_1)}}{M_1} \frac{\Tr{(O_2)}}{M_2} ... \frac{\Tr{(O_k)}}{M_k},
\end{equation}
where $M_n$ is the number of states of the $n^{\mbox{{\scriptsize th}}}$ spin.  To 
compute $\Tr{(O_n)} = \sum_i \left<i\right|O_n\left|i\right>$ we simply use
\begin{eqnarray}
\label{Ipm}
I_{\pm} \left|I, m\right> &=& \sqrt{I (I+1) - m (m \pm 1)} \left|I, m
\pm 1\right>, \\
\label{Iz}
I_{z} \left|I, m\right> &=& m \left|I, m\right>.
\end{eqnarray}

Of course, in order to obtain an algebraic expression for a particular
order of $\tau$, we do not want to consider the sum of all
possible nuclear site indices.  In fact, in order to obtain this
expression, we should not even need to know how many nuclear spins
there are.  Instead, we simply consider all of the possible ``index
configurations.''  An ``index configuration'' indicates
which indices are the same and which are different.  As indicated
above, this is enough information to allow one to evaluate the trace of
the product of spin operators as a rational number [the square roots
in Eq.~(\ref{Ipm}) will always be repeated as factors an even number
of times in the trace].
Additionally, we need only
consider ``index configurations'' that do not obviously result in a
zero trace for any of the indices.  In particular, we only need to
consider configurations in which there are the same number of $I_{+}$
as $I_{-}$ operators for each index, and we can ignore any configuration in
which the only operator for any index is an odd power of $I_z$.

Once the operators and traces are taken care of and we have exhausted
all ``index configurations,'' we are left with terms that have 
factors of $A_n$'s and $b_{nm}$'s, some with the same
nuclear index labels, and some with different labels.  These labels
are to be summed over distinctly (i.e., different labels should
never be assigned the same index).  The labels are
arbitrary, so when checking for ``like'' terms, in order to provide a 
compact expression
for the result, we should consider that the labels can be permuted.  
It would be a waste of computation time to always run through
all permutations of labels when checking to see if two terms are like
terms.  Instead, the problem is transformed into that of checking for graph
isomorphisms, which is a well-studied computational problem.  The program uses
a code called Nauty written by Brendan McKay
(http://cs.anu.edu.au/people/bdm) for checking for graph isomorphisms.

The $A_n$ and $b_{nm}$ factors are transformed into a graph in the
following way.  Each distinct label is represented by a vertex in the
graph.  A factor of $A_i^p$ is represented by coloring the vertex
$i$ to indicate the power $p$.  A factor of $b_{ij}$ is
represented by an edge from vertex $i$ to vertex $j$.  If $b_{ij}$ is
raised to some power (other than 1) then it is represented by making
an edge from both $i$ and $j$ to some special vertex that is colored
in such a way as to indicate that it represents the appropriate power
of $b_{ij}$ (this is just a trick to effectively ``color'' edges).
Then the Nauty program takes this graph and gives a canonical
labeling for the vertices.  If the program comes to another term of
$A_n$ and $b_{nm}$ factors that are equivalent to this one up to a
permutation of labels, Nauty will give it the same canonical labeling
so that the two terms can be combined.

\section{Computation of Tau-Expansion Coefficients}
\label{appendix_tau_computation}

Given an algebraic expression for a particular order, in $\tau$, of
 $v_{E}(\tau)$, using the procedure described in Appendix \ref{appendix_ComputerAlgebra}, 
we need to compute numerical coefficients for
specific applications (with explicit values of $A_{n}$ and
$b_{nm}$).  For a problem that has $N$ nontrivial nuclei and 
an algebraic expression that has $m$ terms and $k$ summation
labels, this will in general take ${\cal O}(m N^{k})$ time to compute.  This
is a huge improvement over direct methods of evolving the state or
diagonalizing the Hamiltonian since there are ${\cal O}[M = (2I + 1)^N]$
states of the system.  So this expansion already saves us from calculations
that grow exponentially with $N$. 
However, when $N$ is on the order of $1000$, which is actually a low estimate
 for considered applications, and we want the ${\cal O}(\tau^{10})$ 
term which has $k=5$, $N^k$ is still quite large ($10^{15}$).
The situation can be drastically improved once more if we are willing
to make another approximation which is very reasonable.  We can set a
threshold for values of $|b_{nm}|$, taking any $|b_{nm}|$ below this
threshold to be zero.  Since $|b_{nm}|$ generally decreases as nuclei
are further separated, this amounts to a near neighbor
approximation (also exploited by our cluster method).  We can then
 adjust this threshold until convergence is achieved.

Let us say that our threshold is set such that, on
average, each nuclei couples to $L$ other nuclei.  
If we assume that
all of the terms of the expression to be calculated are connected, not disjoint,
graphs when representing labels as vertices and factors of $b_{nm}$ as
edges (as discussed in Appendix \ref{appendix_tau_computation} 
in the context of checking for like terms), then this approximation reduces the computation time to
${\cal O}(m N L^{k - 1})$.  For the first label,
there are $N$ possible nuclei to which it can be assigned, but each additional
label must be one of the ${\cal O}(L)$ vertices connected by an edge to a previous vertex/label.

Not all of the terms, in general, have these connected graph
representations, however.  Some of them are disjoint.  This does not
create a real problem though.  These ``disjoint'' terms can be
factored and dealt with separately.  As a simple example,
\begin{eqnarray}
\nonumber
\sum_{i, j, k, l}^{\mbox{{\tiny distinct}}} b_{ij} b_{kl} & = & \sum_{i \ne j} b_{ij} \sum_{k \ne l} b_{kl} - 2 \sum_{i \ne j}
  b_{ij}^2 \\
& & - 4 \sum_{i, j, k}^{\mbox{{\tiny distinct}}} b_{ij} b_{jk},
\end{eqnarray}
where the summation on the left (and one of them on the right) is over distinct indices as indicated above
the summation symbol.  The second term on the right has the factor of
$2$ because it includes $b_{ij} b_{ji} = b_{ij}^2$ (since $b_{nm} = b_{mn}$).  The last term gets a
factor of $2$ from $b_{jk} = b_{kj}$ and another factor of $2$ from 
\begin{equation}
\label{relabel}
\sum_{i, j, k}^{\mbox{{\tiny distinct}}} b_{ij} b_{ik} = 
\sum_{i, j, k}^{\mbox{{\tiny distinct}}} b_{ji} b_{jk} = \sum_{i, j, k}^{\mbox{{\tiny distinct}}} b_{ij} b_{jk},
\end{equation}
in which we relabeled indices.

Our computer program, described in Sec.~\ref{algebra}, automatically performs these conversions as a
preprocess before being fed to the program that performs the numerical
calculations.  In addition to this conversion for ``disjoint'' terms,
the preprocessor performs a simple unidirectionally cascading
factorization, for example, of the form $a [c + d (e + f)]$ but not 
$(a + b) (c + d)$, which reduces
the number of multiplications that must be performed in the
calculation.  This factorization does not change the time complexity of the
calculation, which is ${\cal O}(m N L^{k - 1})$, but it does provide a
noticeable speedup.

\section{Dipolar Perturbation and Cluster Size}
\label{DipolarAndClusterSize}

In this appendix we prove that the contribution of a cluster of size
$k$ is ${\cal O}(\lambda^k)$ with respect to the dipolar perturbation
expansion of Sec.~\ref{perturbation}, at least when 
$\tau \ll \Max{(b_{nm})}^{-1}$.  A direct application of quantum
perturbation theory to the Hahn echo yields Eq.~(\ref{perturb_amp}),
giving a sum of oscillations of amplitude $C_{ijkl} =
F_{ijkl}/D_{ijkl}$ and frequency $\omega_{ijkl} =
\omega_{ijkl}^{(0)} + \lambda \omega_{ijkl}'$.

First, let us only consider the nuclei involved in a term of $F_{ijkl}$
(the numerator of $C_{ijkl}$).  The way in which nuclei ``get
involved'' in $F_{ijkl}$ is via the sum over nuclear pairs in ${\cal
  H}'=\frac{1}{\lambda}{\cal H}^B$.  This gives potentially two nuclei for one $\lambda$, but we
need to do better than that.
Substituting $\left|k_{\pm}\right>$ on the right-hand side of
Eq.~(\ref{eigenvector_recursive}) with the full expression will give successive orders of
$\lambda$.  When one considers a particular term of 
$\left<l^0\right.\left|k_{\pm}\right>$ with the above recursion
formula in mind, one will go from state $\left|l^0\right>$ to state
$\left|k^0\right>$ with some number of intermediate ${\cal H}^0$
eigenstates in between where each state is connected to the next via 
${\cal H}'$.  The second term in the numerator of
 Eq.~(\ref{eigenvector_recursive})
simply corresponds to going from state $\left|l^0\right>$ to
$\left|k^0\right>$ through some number of intermediate states and then
from $\left|k^0\right>$ back to $\left|k^0\right>$ through some more
intermediate states.  The point is simply that one goes from one state
to a different state only via ${\cal H}'$ [Eq.~(\ref{Hpm})].  
With this in mind, Eq.~(\ref{perturb_numer}) indicates that one must go from
state $\left|i^0\right>$ to $\left|j^0\right>$ to $\left|k^0\right>$
to $\left|l^0\right>$ and back to state $\left|i^0\right>$ with some
number of states in between where consecutive states are connected via
${\cal H}'$.  Because we must come back to state $\left|i^0\right>$,
for every lowering operator we pass through, we must also pass through
a raising operator for the same spin.  We now see that we cannot
actually get two nuclei for one $\lambda$; for each
nucleus involved in a given term of $F_{ijkl}$, there is at least one
order of $\lambda$.

The denominator, $D_{ijkl}$, works in a similar way, except that
instead of one cycle from state $\left|i^0\right>$ back to state
$\left|i^0\right>$, one has one of these going from $\left|i^0\right>$ to
$\left|i^0\right>$, another going from $\left|j^0\right>$ to
$\left|j^0\right>$, another going from $\left|k^0\right>$ to
$\left|k^0\right>$, and another going from $\left|l^0\right>$ to
$\left|l^0\right>$.  But the result is the same; because these are all
cyclic, there must be at least one order of $\lambda$ for each nucleus
involved.  One then needs to perform the Taylor expansion of
$D_{ijkl}^{-1}$.  Let $g(\lambda) = D_{ijkl}(\lambda)$ and
$f(\lambda) = \frac{1}{g(\lambda)}$.  The
Taylor series expansion is
\begin{equation}
\label{taylor}
f(\lambda) = \sum_{n=0}^{\infty} f^{(n)}(0)\frac{\lambda^n}{n!},
\end{equation}
where $f^{(n)}(\lambda)$ may be obtained recursively with
\begin{equation}
\label{g_derivatives}
\left(\frac{g^{(m)}(\lambda)}{(g(\lambda))^k}\right)' 
= \frac{g^{(m+1)}(\lambda)}{(g(\lambda))^{k}} - k \frac{g^{(m)}(\lambda)}{(g(\lambda))^{k+1}},
\end{equation}
and then we note that $[g(0)]^k = 1$ so there will not be anything in
the denominators of Eq.~(\ref{g_derivatives}) to worry about in the
context of Eq.~(\ref{taylor}).  Examining the numerators in
Eq.~(\ref{g_derivatives}), we note that for each derivative of $f(\lambda)$, 
we will lose at most one order of
$\lambda$.  But in Eq.~(\ref{taylor}) for each derivative there is a
factor of $\lambda$ provided to compensate.  If we write the $\lambda$ 
expansion of $g(\lambda)$ as
\begin{equation}
g(\lambda) = \sum_{n=0}^{\infty} g_n \lambda^n,
\end{equation}
then, when we make the transformation to $f(\lambda) = 1/g(\lambda)$,
each instance of $g_n$ will be accompanied by at least $n$ orders of $\lambda$.
Therefore, after Taylor expanding the inverse of $D_{ijkl}$, we
preserve the property that there are at least as many orders of
$\lambda$ as the number of nuclei involved in a given term.

We have dealt with the numerator and denominator of the amplitude, but
we must still consider the frequency.  We can rewrite the
exponential in Eq.~(\ref{perturb_amp}) as
\begin{eqnarray}
e^{-\imaginary \omega_{ijkl} \tau}&=&
e^{-\imaginary \omega_{ijkl}^{(0)} \tau}~
\exp{\left(-\imaginary \lambda \omega_{ijkl}' \tau\right)} \\
\label{perturb_freq_expansion}
&=&e^{-\imaginary \omega_{ijkl}^{(0)} \tau}
\sum_{n=0}^{\infty} \frac{\left(-\imaginary \lambda \omega_{ijkl}' \tau\right)^n}{n!}.
\end{eqnarray}
Thus we have transformed the perturbative part of the frequency into
an amplitude (with a time dependence as well).  We can now use a similar
argument that was used for $F_{ijkl}$ and $D_{ijkl}$.  Each term of
Eq.~(\ref{omega'}) will cycle from some state back to the same state
with ${\cal H}'$ connecting each intermediate state.  As a result,
there must be at least as many orders of $\lambda$ in a given term
of $\lambda \omega_{ijkl}'$ as there are nuclei involved.  There will, 
however, be one extra $b_{nm}$-type factor without an accompanying
$1/(A_n - A_m)$.  This gets combined with $\tau$ and will be small as
long as $\Max{(b_{nm})} \tau \ll 1$.  All of this can be exponentiated
in the series of Eq.~(\ref{perturb_freq_expansion}), but that can only
increase orders of $\lambda$ without increasing the number of nuclei involved.

There is one final consideration.  How many nuclei are involved
in $\exp{\left(-\imaginary \omega_{ijkl}^{(0)} \tau\right)}$?
From Eqs.~(\ref{omega0}) and (\ref{E_0}), we see that this will include
the nuclei that differ between states
$\left|i^0\right>$ and $\left|j^0\right>$ along with the nuclei that
differ between states $\left|k^0\right>$ and $\left|l^0\right>$.  Since
a term of $F_{ijkl}$ will require linking state $\left|i^0\right>$ to 
$\left|j^0\right>$ to $\left|k^0\right>$ to $\left|l^0\right>$ and
back to $\left|i^0\right>$ via ${\cal H}'$ (through some number of
nuclear states), it must involve at least that many nuclei.
So this factor will
neither increase the number of involved nuclei nor change the number
of $\lambda$ factors.

We therefore conclude that, using an approximation that assumes 
$\Max{(b_{nm})} \tau \ll 1$, 
there will be at least as many orders of $\lambda$
in a term of the $\lambda$-expansion of Eq.~(\ref{perturb_amp})
 as there are nuclei involved.  
Therefore a cluster contribution of size $k$ will be of
${\cal O}(\lambda^k)$ or a higher order of $\lambda$.  This justifies the
cluster expansion in the regime in which $c_{nm} \gg 1$ and 
$\Max{(b_{nm})} \tau \ll 1$.
To be more accurate and to relax the $\Max{(b_{nm})} \tau \ll
1$ constraint, as long as $\Max{(b_{nm})} \tau < 1$ we have shown that
clusters of size $k$ will give a contribution of order
$\lambda^{k-1}$.  Because it is consistent for different cluster
sizes, $k$, and because clusters of size one do not contribute,
this reduction by an order in $\lambda$ does not
have an impact upon the validity of the cluster expansion in any of
its various forms (i.e., product form or exponentiated form).  For
simplicity of the discussion, however, we will regard $\Max{(b_{nm})}
\tau$ simply as another order of $\lambda$.

%
%


\begin{thebibliography}{99}
\bibitem{footnote1}
It is possible for virtual spin-flip transition between
hyperfine-coupled electron and nuclear spins to produce an effective
hyperfine-mediated interaction between distant nuclei.  Although it can be
suppressed by increasing the external magnetic field strength, it can 
be an important effect for free induction decay
(Refs.~\onlinecite{deng} and \onlinecite{yao05}); 
however, under the application of refocusing pulses, this leads only to
a small visibility decay (Ref.~\onlinecite{shenvi05}).
\bibitem{footnote2}
Our interest is in the ``intrinsic'' decoherence in the context of
refocused spin echoes which is dominated by spectral diffusion.  
Free induction decay (without refocusing) is important during the
duration of a single gate operation.  Hyperfine-induced free induction
decay has been recently studied in literature (Refs.~\onlinecite{klauser, deng}).
\end{thebibliography}
\end{document}